\PassOptionsToPackage{dvipsnames}{xcolor}
\documentclass[longauth,traditabstract]{aa}  

\usepackage{graphicx}
\usepackage{txfonts}
\usepackage{soul}
\usepackage{placeins}
\usepackage{booktabs} 
\usepackage{multirow}
\usepackage[normalem]{ulem}
\usepackage[colorlinks=true,citecolor=blue,linkcolor=blue,urlcolor=blue, pdftex, pdfauthor={Justyn Campbell-White}]{hyperref}
\newcommand{\teff}{$T_{\rm eff}$}
\newcommand{\logg}{$\log g$}
\newcommand{\vsini}{$v\sin i$}

\newcommand{\leqsim}{\,\raisebox{-0.6ex}{$\buildrel < \over \sim$}\,}
\newcommand{\geqsim}{\,\raisebox{-0.6ex}{$\buildrel > \over \sim$}\,}

\usepackage[dvipsnames]{xcolor}

\begin{document} 

\title{Empirical Determination of the Lithium 6707.856\,\AA\ Wavelength \\in Young Stars \thanks{Based on observations collected at the European Southern Observatory under ESO programmes 105.207T and 106.20Z8.}}

\author{Justyn Campbell-White\inst{\ref{instESO},\ref{instUOD}}
        \and
        Carlo F. Manara\inst{\ref{instESO}}
        \and
        Aurora Sicilia-Aguilar \inst{\ref{instUOD}}
        \and
        Antonio Frasca \inst{\ref{instOACT}}
        \and
        Louise D. Nielsen\inst{\ref{instESO}}
        \and \\
        P. Christian Schneider\inst{\ref{instHAM}}
        \and 
        Brunella Nisini\inst{\ref{instOROM}}
        \and
        Amelia Bayo\inst{\ref{instESO},\ref{instVALP}}
        \and
        Barbara Ercolano\inst{\ref{instLMU},\ref{instECO}}
        \and 
        P\'eter \'Abrah\'am\inst{\ref{instAK1},\ref{instAK2},\ref{instAK3}}
        \and
        Rik Claes\inst{\ref{instESO}}
        \and \\
        Min Fang\inst{\ref{instPM}}
        \and
        Davide Fedele\inst{\ref{instFIR}}
        \and
        Jorge Filipe Gameiro\inst{\ref{instPOR1},\ref{instPOR2}}
        \and
        Manuele Gangi\inst{\ref{instASI},\ref{instOROM}}
        \and
        \'Agnes K\'osp\'al\inst{\ref{instAK1},\ref{instAK2},\ref{instAK3},\ref{instAK4}}
        \and\\
        Karina Mauc\'{o}\inst{\ref{instESO},\ref{instUNAM}}
        \and
        Monika G. Petr-Gotzens\inst{\ref{instESO}}
        \and 
        Elisabetta Rigliaco\inst{\ref{instPAD}}
        \and
        Connor Robinson\inst{\ref{instAMH}}
        \and
        Michal Siwak\inst{\ref{instAK1},\ref{instAK2}}
        \and\\
        Lukasz Tychoniec\inst{\ref{instESO}}
        \and 
        Laura Venuti\inst{\ref{instSETI}}
      }

\institute{European Southern Observatory, Karl-Schwarzschild-Strasse 2, 85748 Garching bei M\"unchen, Germany\label{instESO}\\
          \email{jcampbel@eso.org}
\and
SUPA, School of Science and Engineering, University of Dundee, Nethergate, Dundee DD1 4HN, U.K.\label{instUOD}
\and
INAF - Osservatorio Astrofisico di Catania, via S. Sofia, 78, 95123 Catania, Italy\label{instOACT}
\and
Hamburg Observatory, Gojenbergsweg 11, 21029 Hamburg, Germany\label{instHAM}
\and
INAF - Osservatorio Astronomico di Roma: Monte Porzio Catone, Lazio, Italy\label{instOROM}
\and
Instituto de F\'{\i}sica y Astronom\'{\i}a, Universidad de Valpara\'{\i}so, Chile\label{instVALP}
\and
University Observatory, Faculty of Physics, Ludwig-Maximilians-Universität München, Scheinerstr. 1, 81679 Munich, Germany\label{instLMU}
\and
Excellence Cluster 'Origins', Boltzmannstr. 2, 85748 Garching, Germany\label{instECO}
\and
Konkoly Observatory, Research Centre for Astronomy and Earth Sciences, E\"otv\"os Lor\'and Research Network (ELKH), Konkoly-Thege Mikl\'os \'ut 15-17, 1121 Budapest, Hungary\label{instAK1}
\and
CSFK, MTA Centre of Excellence, Konkoly-Thege Mikl\'os \'ut 15-17, 1121 Budapest, Hungary\label{instAK2}
\and
ELTE E\"otv\"os Lor\'and University, Institute of Physics, P\'azm\'any P\'eter s\'et\'any 1/A, 1117 Budapest, Hungary\label{instAK3}
\and
Purple Mountain Observatory, Chinese Academy of Sciences, 10 Yuanhua Road, Nanjing 210023, China\label{instPM}
\and
INAF-Osservatorio Astrofisico di Arcetri, L.go E. Fermi 5, I-50125 Firenze, Italy\label{instFIR}
\and
Instituto de Astrof\'{i}sica e Ci\^{e}ncias do Espa\c{c}o,  Universidade do Porto, CAUP, Rua das Estrelas, 4150-762 Porto, Portugal\label{instPOR1}
\and
Departamento de F\'{\i}sica e Astronomia, Faculdade de Ci\^encias, Universidade do Porto, rua do Campo Alegre 687, 4169-007 Porto. Portugal\label{instPOR2}
\and
ASI, Italian Space Agency, Via del Politecnico snc, 00133 Rome, Italy\label{instASI}
\and
Max Planck Institute for Astronomy, K\"onigstuhl 17, 69117 Heidelberg, Germany\label{instAK4}
\and
Instituto de Astronom\,ia, Universidad Nacional Autónoma de M\'exico, Carr. Tijuana-Ensenada km107, C.I.C.E.S.E., 22860 Ensenada, B.C., M\'exico\label{instUNAM}
\and
INAF - Osservatorio Astronomico di Padova, Vicolo dell'osservatorio 5, 35122 Padova, Italy\label{instPAD}
\and
Department of Physics \& Astronomy, Amherst College, C025 Science Center 25 East Drive, Amherst, MA 01002, USA\label{instAMH}
\and
SETI Institute, 339 Bernardo Ave, Suite 200, Mountain View, CA 94043, USA\label{instSETI}
}


\titlerunning{Lithium 6707.856\,\AA\ Measured in YSOs}

 
  \abstract
 {
 Absorption features in stellar atmospheres are often used to calibrate photocentric velocities for kinematic analysis of further spectral lines. The Li feature at $\sim$6708\,\AA\ is commonly used, especially in the case of young stellar objects for which it is one of the strongest absorption lines. However, this is a complex line comprising two isotope fine-structure doublets.
 We empirically measure the wavelength of this Li feature in a sample of young stars from the PENELLOPE/VLT programme (using X-Shooter, UVES and ESPRESSO data) as well as HARPS data. For 51 targets, we fit 314 individual spectra using the STAR-MELT package, resulting in 241 accurately fitted Li features, given the automated goodness-of-fit threshold.
 We find the mean air wavelength to be 6707.856\,\AA, with a standard error of 0.002\,\AA\ (0.09\,km/s) and a  weighted standard deviation of 0.026\,\AA\ (1.16\,km/s). The observed spread in measured positions spans 0.145\,\AA, or 6.5\,km/s, which is up to a factor of six higher than typically reported velocity errors for high-resolution studies. 
 We also find a correlation between the effective temperature of the star and the wavelength of the central absorption.
 We discuss how exclusively using this Li feature as a reference for photocentric velocity in young stars could potentially be introducing a systematic positive offset in wavelength to measurements of further spectral lines. If outflow tracing forbidden lines, such as [\ion{O}{i}]\,6300\,\AA, are actually more blueshifted than previously thought, this then favours a disk wind as the origin for such emission in young stars.
 }
   

   \maketitle
%
\section{Introduction}
Lithium is a benchmark element in stellar astrophysics and causes prominent photospheric features in optical spectra of young stellar objects (YSOs). In particular, the \ion{Li}{i}\,$\sim$6708\,\AA\ feature is often (one of) the strongest photospheric absorption lines; with lines from other metals being more prone to line-dependent veiling. This Li feature is used to age star clusters from the lithium depletion boundary \citep[e.g.][]{messina_rotation-lithium_2016,gutierrez_albarran_gaia-eso_2020,binks_gaia-eso_2022} and to identify young stellar membership  \citep[e.g.][]{walter_x-ray_1994,montes_chromospheric_2001,bayo_spectroscopy_2011,jeffries_gaia-eso_2017}. Furthermore, the position  of the Li feature is fundamentally important as a reference for establishing the photocentric velocity, in particular for kinematic studies of YSO emission line properties.

The absorption feature consists of multiple fine\footnote{and hyper-fine structure transitions, which will not be considered here} structure transitions of both $^7$Li and $^6$Li \citep{kurucz_primordial_1995,morton_atomic_2003}. The typical abundance of the more fragile $^6$Li has been found to be <10\% for a range of stellar types \citep[e.g.][and references therein]{fields_implications_2022}. The complexity of this absorption feature is due to the two spin-orbit electronic transitions from the 2P to the 2S level of $^7$Li, alongside any lesser contribution from the same transitions of $^6$Li. Table\,\ref{tab:liwl} shows the air wavelengths of these components (we will use air wavelengths throughout). From the L-S coupling and the gyromagnetic ratios, each doublet is in a 2:1 resonance, with the $D_2$ lines the stronger of the pairs. Additionally, these lines are adjacent to spectral lines of \ion{Fe}{i}, \ion{Si}{i}, \ion{V}{i}, and \ion{Cr}{i} \citep[see Table 2 in][and Fig.\,\ref{fig:li_spec}]{franciosini_gaia-eso_2022}, which commonly appear blended with the Li feature. From this simple consideration, it is clear that the centroid of this feature should not have a constant position in YSO spectra.

Previous studies that took the Li feature as reference for photocentric calibration of YSO spectra have used positions from 6707.800\,\AA\ to 6707.876\,\AA\ \citep{edwards_forbidden_1987,cabrit_forbidden-line_1990,natta_x-shooter_2014,pascucci_narrow_2015,simon_tracing_2016,nisini_connection_2018}. This is performed so that further spectral features, such as outflow tracing forbidden lines, can be kinematically measured with respect to the stellar photospheric rest frame. A commonly used emission line is [\ion{O}{i}]\,6300\,\AA, which can be readily decomposed into a high- ($v\geqsim $30--50 km/s) and low- velocity component (HVC, LVC, respectively); with the LVC further comprised of a narrow- (NC) and broad-component (BC) \citep{banzatti_kinematic_2019,pascucci_role_2022}. Small blueshifts of the LVC centroid by a few km/s were noted by e.g. \citet{hartigan_disk_1995} and \citet{rigliaco_understanding_2013}, who suggest this is a clear sign of unbound gas flowing outwards from the disk. \citet{banzatti_kinematic_2019} found that for low LVC blueshifts from disks possessing an inner cavity, the blueshift reduces as the inner dust from the disk clears and emission originates from further out in the disk. However, surveys of the [\ion{O}{i}]\,6300\,\AA\ line in YSOs show that this line is centered at the photocentric velocity for almost half of the samples \citep[][Nisini et al. in prep.]{simon_tracing_2016} with another possible source of the LVC being gas bound to the star/inner-disk. It is still difficult to disentangle the possible sources of the forbidden line emission, be it photoevaporative, magnetohydrodynamic (MHD), or non-thermal outflows \citep{nemer_role_2020}, especially for the NC of the LVC. \citet{weber_interpretation_2020} showed that photoevaporation reproduces most of the observed NCs but not the BCs or the HVCs, whereas MHD winds are able to reproduce all components but produce Keplerian double peaks that do not agree with observations. It is therefore important to determine whether or not the LVC is centred at the photocentric rest velocity, or shifted, with models from \citet{ercolano_blueshifted_2016} suggesting that even the slightest blue-shift is a tell-tale sign of a disk wind. In  massive environments, external photoevaporation could lead to shifts of less than $\sim$2 km/s \citep{ballabio_oi_2022}.

Since these [\ion{O}{i}] 6300\,\AA\ and other forbidden line LVC velocities are measured within a few km/s of the stellar velocity, it is clear that accurate determination of such velocity for YSOs is critical for subsequent kinematic analysis. 
In this letter, we report a systematic, empirical study of the position of the Li $\sim$6708\,\AA\ feature, from a sample of YSOs using mid- to high-resolution spectrographs. This will allow future studies of forbidden line kinematics to avoid potential systematic offsets that were, until now, unforeseen.

\begin{table}
\centering
\caption{Line centers and oscillator strengths of the components of the \ion{Li}{i}\,$\lambda$6708\,\AA\ line \citep{morton_atomic_2003}}
\label{tab:liwl}
\begin{tabular}{c|l|c|c}

 Li Isotope & Fine Structure & Air Wavelength [\AA] & $f$\\
  \hline
  $^7$Li D$_1$ & (2 $^{2}$P$_{1/2}$ - 2 $^{2}$S$_{1/2}$) & 6707.9147 & 0.249\\
  $^7$Li D$_2$ & (2 $^{2}$P$_{3/2}$ - 2 $^{2}$S$_{1/2}$) & 6707.7637 & 0.498\\
  $^6$Li D$_1$ & (2 $^{2}$P$_{1/2}$ - 2 $^{2}$S$_{1/2}$) & 6708.0728 & 0.249\\
  $^6$Li D$_2$ & (2 $^{2}$P$_{3/2}$ - 2 $^{2}$S$_{1/2}$) & 6707.9219 & 0.498
\end{tabular}
\end{table}


\section{YSO sample and observations}

The YSO spectra used in this work were obtained using the ESO instruments ESPRESSO, UVES and X-Shooter (on the ESO VLT), and HARPS (on the ESO 3.6\,m). VLT data are from the public PENELLOPE survey \citep{manara_penellope_2021}. This was a contemporaneous survey to the \textit{HST}/ULLYSES program\footnote{Part of the ODYSSEUS collaboration (Outflows and Disks around Young Stars: Synergies for the Exploration of ULLYSES Spectra, \citep{espaillat_odysseus_2022} \url{https://sites.bu.edu/odysseus/}}, during which 82 young stars were observed by \textit{HST}.  Spectroscopic data reduction was carried out using the ESO Reflex workflow v2.8.5 \citep{freudling_automated_2013}. Telluric correction was also performed using {\sc Molecfit} \citep{smette_molecfit_2015,kausch_molecfit_2015}.\footnote{For further details of the PENELLOPE observation strategy and data reduction, see \citet{manara_penellope_2021}.}

Here, we include repeated VLT observations of 41 stars from the ULLYSES sample that are from the Orion OB1 and $\sigma$-Orionis associations \cite[as presented in][]{manara_penellope_2021} as well as Cha I, $\eta$ Cha and Lupus. This sample covers YSOs with spectral types K2 through M5, and a wide range of disk ages ($\sim$1--10\,Myr) and evolutionary stages. 

We also include ten YSOs with repeated HARPS observations from Pr. ID. 105.207T (P.I. Campbell-White), acquired to investigate the temporal spectral evolution of YSOs of different mass ranges. This sub-sample includes targets at earlier spectral type (G5 through M1) than the PENELLOPE sample. The HARPS data for four targets also have more repeated observation of each star (from eight to 15, see Appendix\,\ref{app:individual_results}). The 1D spectra were used, with data reduction carried out using the standard ESO pipeline and no further corrections applied. There is no significant influence from telluric lines around the position of the Li 6708\,\AA.

No photospheric corrections are applied to any of the data used in this work. Although this could potentially remove blends from around the Li absorption (if using a Li-poor template or a synthetic spectrum), with the aim of this investigation being to understand the measured position of the Li absorption, we do not consider such correction. 
Moreover, as the centroid of the Li feature is normally measured in the observed spectra without subtracting a photospheric template, we have purposely made our measurements including the effect of the line blends.


\section{Methodology}

From the observations of the 51 targets outlined, 314 YSO spectra were fitted with STAR-MELT: 92 from HARPS, 72 from UVES, 61 from X-Shooter and 88 from ESPRESSO. Barycentric correction was applied to the UVES and X-Shooter air wavelength frames, ESPRESSO and HARPS already use this standard. Stellar photocentric radial velocities (RVs) were calculated for each individual observation around the wavelength range of 6000-6250\,\AA, using the standard RV templates of the STAR-MELT package \citep{campbell-white_star-melt_2021}. This ensured that the measurement of the Li centroid was with respect to any intrinsic stellar RV variations. We selected this range to include further prominent photospheric absorption lines \citep[as in e.g.,][]{Fang2018} close to but not including the Li absorption line itself to avoid wavelength-dependent RV variations, \citep[e.g.,][]{martin_multiwavelength_2006}. 

An arbitrary reference wavelength of 6708.0\,\AA\ was then used to measure positions of the Li absorption lines, which were fitted with a single Gaussian plus a linear component (the latter to fit the local continuum and help account for asymmetries). A velocity range of 150\,km/s was used for each fit. This adequately covered the full absorption plus continuum of each Li observation. The centre of the Gaussian component was obtained for each observation, with associated standard errors from the least-squares optimisation fitting. The shift with respect to the reference wavelength thus gives the centroid/photocenter of the Li absorption line for each observation.

The automated fitting and goodness-of-fit restrictions using STAR-MELT (see Appendix\,\ref{app:fits_errors}) resulted in 241 accurately fitted individual observations across the YSOs spectra, over 75\% of the initial sample. Reasons for the subset of larger errors include, primarily, lower S/N data; binaries \citep[e.g. CVSO104;][]{frasca_penellope_2021}; and non-detections of Li from the HARPS sample, also due to low S/N, likely continuum veiling and/or stellar age. Li was detected in all YSOs from the PENELLOPE sample. 

We note that in the recent work of \citet{franciosini_gaia-eso_2022} on Li abundances from the Gaia/ESO survey multiple Gaussian component fits were used for the UVES spectra to model both components of the Li doublet (for the FGK stars), as well as adjacent Fe and Si blends. Whilst this is important for measuring equivalent widths (EWs), we find that for our YSO sample from UVES (and even the higher resolution ESPRESSO and HARPS data), the feature centre is well-fitted by the single Gaussian plus linear component in each instance. Increasing the number of components in an attempt to more accurately model the doublet does not improve the calculated $\chi^2$ goodness-of-fit, nor shift the overall centre of the fit by more than the typical measurement errors we report ($\leqsim1$ km/s, see Fig\,\ref{fig:li_fits}). Since we are interested in the position of the feature centroid (as it is measured for photocentric RV calibrations), the fitting procedure we use remains accurate in this sense.

\begin{figure}[]
   \includegraphics[width=0.5\textwidth]{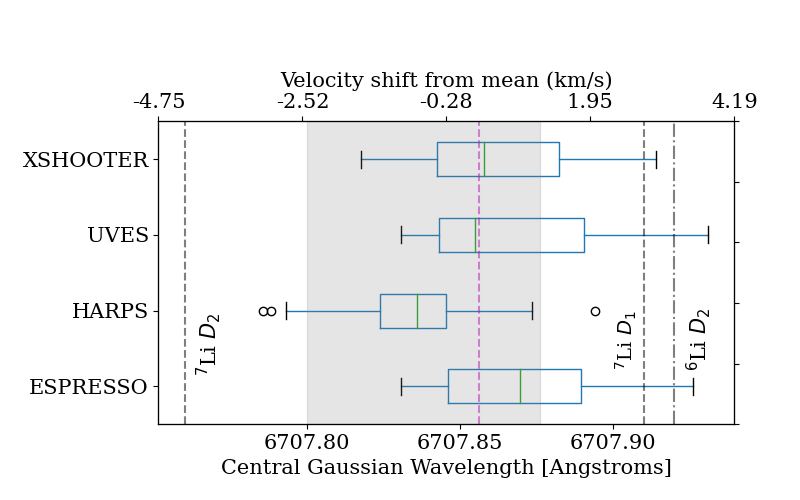}
\caption{Box plot of resulting central wavelength position from fitting all of the PENELLOPE VLT (ESPRESSO, UVES, X-Shooter) and the further HARPS YSO spectra. The boxes extend from the lower to upper quartile values of the data, with a green line at the median. The whiskers extend to 1.5 times the box range, with outliers indicated by circles. Mean central wavelength position is marked by the magenta dashed line. Positions of the $^7$Li $D_1$, $D_2$ and $^6$Li $D_2$ lines are indicated by the black dashed and dashed-dot lines.  The grey shaded area indicates the range of values of the \ion{Li}{i} line centres used as reference for RV calculations in previous works. }
\label{fig:li_results}
\end{figure}


\section{Analysis \& discussion}

\begin{figure*}[]
\centering
   \includegraphics[width=0.8\textwidth]{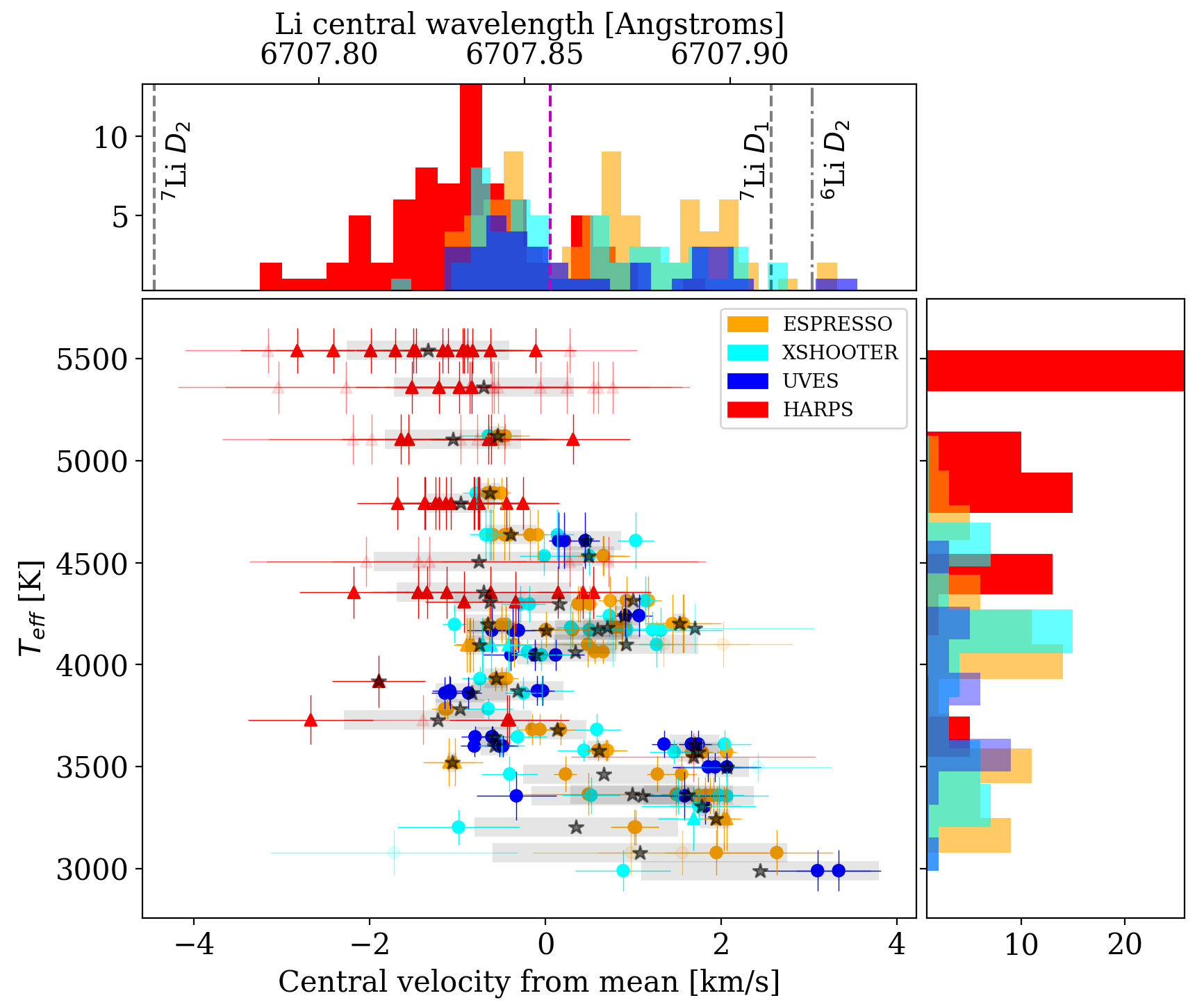}
\caption{\teff versus velocity difference from mean measured Li centroid position. Stars indicate the mean measured position for each target, with the shaded boxes showing the 1$\sigma$ spread of measurements. Circles indicate that \teff\ measurements are from ROFIT/PENELLOPE spectra, and triangles are taken from the TESS Input Catalogue. We note an anti-correlation (r=$-$0.58) between \teff\ and mean position, with lower \teff\ stars showing more redshifted central values. Total velocity error bars are from adding the standard errors of the RV and Gaussian centre measurements in quadrature. Faded markers and error bars are measurements with total velocity errors > 0.75 km/s. Mean measured Li centroid and positions of the $D$ line components are indicated on the top histogram.}
\label{fig:teff}
\end{figure*}

Figure\,\ref{fig:li_results} shows a box plot of the measured Li positions across all data, split by instrument. The overall mean wavelength of the Li absorption line is 6707.856\,\AA, with a standard error of 0.002\,\AA\ (0.09\,km/s) and a weighted standard deviation of 0.026\,\AA\ (1.16\,km/s). This observed spread in measured values spans 0.145\,\AA, or 6.5\,km/s. The grey shaded area on Fig.\,\ref{fig:li_results} indicates the range of Li wavelengths used to measure stellar RVs from a sample of previous studies \citep{natta_x-shooter_2014,pascucci_narrow_2015,simon_tracing_2016,nisini_connection_2018}.

\subsection{The wavelength of Li absorption}

The Li absorption feature mean wavelength we observe is in good agreement with previous studies to measure this line \citep[e.g. 6707.851 $\pm$ 0.007\,\AA, from observations of a carbon star][]{wallerstein_ratio_1977}. Figure\,\ref{fig:li_results} shows that the median values obtained from the UVES and X-Shooter data are also close to this overall value. The ESPRESSO and HARPS data, however, possess median values either side of the overall sample mean. Since these instruments are of similar resolving power, we believe this is an observed effect from the sub-samples, with any potential instrumental influences carefully checked. In the following subsection, we discuss the most prominent correlation with respect to the observed spread of Li centroids. To preface this, we first briefly discuss further potential sources of the Li shifts. 

For YSOs, accretion-related activity causes wavelength-dependent continuum veiling. Corrections for veiling have been made to the observed Li feature in T Tauri stars \citep[e.g.][]{1991A&A...252..625B,biazzo_x-shooter_2017}, and determination of the photospheric abundances using NLTE effects in YSOs indicate that they usually have close to cosmic lithium abundance \citep{1992ApJ...392..159M,1994ApJ...436..262M}. It is likely that the Li$^6$/Li$^7$ isotopic ratio of YSOs is also cosmic. Recently, \citet{2022MNRAS.509.1521W} used 3D NLTE radiative transfer methods to model ESPRESSSO spectra of metal poor stars, finding extremely low isotopic Li$^6$/Li$^7$ ratios, however, this has not yet been carried out for YSOs. We hence assume that differences in the Li isotopic ratio is negligible and not significantly influencing the observed Li feature position in our sample of young stars. 

Slight differences in the lithium abundance across different stars may, however, affect the observed position by determining how influential the adjacent \ion{Fe}{i} line blend is on the overall shape and centre of the feature. If the \ion{Fe}{i} is stronger, then we would expect a larger blueshift of the photocentre of the whole feature. Metallicity of the stars would also contribute significantly \citep{franciosini_gaia-eso_2022}. For cool, active stars, the fractional coverage of star-spots also affects the measured Li abundances \citep{franciosini_gaia_2022}, which is further discussed in this section. 

We find no clear correlation between the EW of the Li feature and its central wavelength, suggesting that the intensity of the \ion{Li}{i} plays a minor role in its observed position, at least in the range of abundances covered by our sample. Although it may be that the lines more blended have larger EWs, as the Li EW increases with the elemental abundance, any potential trend here is not apparent. We obtain similar results as those from \citet{biazzo_x-shooter_2017}, who find EWs and \teff\ in good agreement with NLTE curves of growth, however, we do not see exclusively the lowest \teff\ stars showing the smallest EWs, suggesting our sample have higher overall abundances. 

As outlined, another potential contributor to the spread of Li position values is due to the active nature of YSOs. Indeed, even in the case of non-accreting objects, the strong magnetic activity gives rise to cool photospheric starspots which can distort the absorption line profiles and change the centroid \citep[e.g.,][]{biazzo_young_2009,lanza_gaps_2018}. Furthermore, accretion-related activity can affect the measured RVs \citep{sicilia-aguilar_accretion_2015,alcala_x-shooter_2017,campbell-white_star-melt_2021}. Continuum veiling differences could have an impact on the measured abundances and EWs, both of the Li and of the adjacent lines. We see from ROTFIT analysis of the PENELLOPE data that the veiling changes daily \citep{manara_penellope_2021}. Recent work on RU Lup by \citet{stock_accretion_2022} shows significant veiling differences for this K7 star across 12 nights. We may also have line-dependent veiling such that the Li remains in absorption, but the blended Fe or Si lines could have an emission component, thus altering the measured centroid of the absorption line. From our results, we detect no significant correlation with the veiling measured at 650\,nm and central Li position found (see Fig.\,\ref{fig:veil_comp}). Therefore, if veiling does have an influence, it appears to not be systematically affecting the position of the Li line. We also find no significant correlation between the \logg\ of the YSOs and the central position of the Li (Fig.\,\ref{fig:logg_comp}). 

\subsection{Effective temperature correlation}

Figure\,\ref{fig:teff} shows the measured central position of the Li feature versus the effective temperature (\teff) of the star. Points are coloured by instrument, with histograms showing the two distributions. In the top panel, the corresponding measured Li wavelength is also shown, along with the positions of the $^7$Li $D_1$ and $D_2$ and the $^6$Li $D_1$ components. The mean value is marked by the magenta line. 

\teff\ values for the PENELLOPE data are calculated with ROTFIT \citep{frasca_further_2003,frasca_x-shooter_2017}. \teff\ values for the HARPS sample are from the \textit{TESS} Input Catalogue \citep[TIC v8.2,][]{stassun_revised_2019}. TIC temperatures are either compiled from the literature, to favour spectroscopic measurements, or determined using dereddened photometric estimates that they compute from \textit{Gaia} bands. They quote a low deviation between photometric and spectroscopic measurements for their entire sample (median 10\,K), however, we find when comparing common values from the PENELLOPE sample, larger offsets; in accordance to those reported by \citet{gangi_giarps_2022} and \citet{flores_effects_2022} (see Fig.\,\ref{fig:teff_comp}). The HARPS \teff\ values may hence be up to a few 100\,K lower when accounting for spot coverage using spectroscopic determinations. 

The observed spread in Li positions thus appears to be in part due to the \teff\ of the stars. We find a Pearson correlation coefficient of $r = -0.58$ ($p = 1\times 10^{-22}$) for the entire sample ($p = 9\times 10^{-5}$ for the mean position values for each target). A corresponding linear fit relationship is $ \lambda = $ \teff\ $ \times\ (-2.206\times 10^{-5} \pm 2\times 10^{-6}) + (6707.950 \pm 0.009) $. From the box-plots in Fig.\,\ref{fig:li_results} and the histograms in Fig.\,\ref{fig:teff}, we see that although the UVES and ESPRESSO Li positions cover the same range, the mean Li position is higher for the ESPRESSO data; which can be explained by the ESPRESSO sub-sample containing slightly lower \teff\ stars. The HARPS sample clearly includes stars with the highest \teff. Even if these points were moved down by a few 100k (as discussed above), the correlation would still be statistically significant. The X-Shooter results (which are all from repeated observations of either ESPRESSO or UVES targets) show a lower minimum of Li positions, although with almost the same mean as the UVES and overall sample.

\subsection{Spread from individual stars}

Results from repeated observations show that the position of the Li feature changes with time. Since each of the position measurements are in the photocentric frame, this is not due to any orbital motion, with known binaries excluded. This may be due to convective layers recycling material that can contribute to the blend. It is more likely due to the star-spot coverage distorting the shape of the absorption lines and/or short term variations in veiling. Since we have more repeated observations from the HARPS data, it may appear from Fig\,\ref{fig:teff} that the highest dispersion is just from these observations, however, we note that also in the VLT repeated observations, in some cases, we observe a larger dispersion. Table\,\ref{tab:positions} shows the summary of observations of each target, with the mean RVs and Li position shifts and associated standard deviations for the repeated observations. The majority of these repeated observations show a 1$\sigma$ spread of < 0.5 km/s for the Li central position. The RV spread is slightly higher, with these two spreads not correlated. 

For certain stars, e.g. BP Tau with X-Shooter, we see a large variance in the RV values, but a low spread in measured Li feature centre, suggesting that although the observed RV changes, the position of the Li feature is more stable. Furthermore, for the lower resolution X-Shooter observations, the standard calibrations result in a stated RV accuracy of 7.5\,km/s, with determined values approaching this for stars with higher \vsini\ or lower S/N \citep{frasca_x-shooter_2017}. Conversely, for RXJ10053-7749 and MS Lup with HARPS, we see a low RV variation between observations, but a larger spread in Li wavelength values. These type of RV spreads are typical for YSOs and are due to the accretion and activity affecting the measured results. They are, however, less than the threshold for spectroscopic binary detection \citep[e.g.][]{almeida_finding_2012}. 

For the four stars with the most repeated observations, with coverages exceeding their rotational periods, we find that the temporal variation of the Li centroids may be on shorter timescales than the rotation periods (see Appendix\,\ref{app:individual_results}). This may be further evidence of multiple star spots influencing the shift of the Li feature. It may also be due to the variability in veiling affecting the measured position, which has recently been shown to not vary periodically with stellar rotation \citep{2023A&A...670A.142S}. Furthermore, these results suggest that the range in Li position values we obtain for targets with fewer observations may not cover the extent of the positional variability. However, the reported mean position values should still be a good representation of the overall trend.

\subsection{Implications for kinematic studies of [\ion{O}{i}]}

For outflow tracing emission lines from YSOs, such as [\ion{O}{i}], kinematic studies are useful to infer the potential origin of the emission, be it from a disk wind (from an MHD or photoevaporative origin) or rather from gas bound to the star/disk system at a low density. These scenarios are regarding the LVC of the emission, with the HVC accepted to trace the collimated, high-velocity jet \citep{hartigan_disk_1995,banzatti_kinematic_2019}. 

For an example case of TW Hya, the [\ion{O}{i}] 6300\,\AA\ line is centred on the photocentric rest frame, suggesting that either the emission originates from the innermost parts of the disk, bound to the star-disk system, possibly due to dissociation of OH molecules within the disk; or that the emission originates from the inner cavity of the dust-cleared disk \citep{pascucci_photoevaporative_2011}. For the former scenario, \citet{gorti_emission_2011} showed that the [\ion{O}{i}] line ratios are in agreement with that expected by stellar FUV photons causing the dissociations, which was observationally favoured by \citet{rigliaco_understanding_2013}, but generally, this requires a large FUV intensity. For the latter, \citet{banzatti_kinematic_2019} showed that further YSOs hosting a disk with a cavity had blueshifted [\ion{O}{i}] LVCs, that approach the stellar velocity as the size of the cavity increases, which had previously been demonstrated in the models of \citet{ercolano2010}.

If the Li feature alone is used for calibrating the photocentric RV, from the results we present here, we see that potential systematic offsets for the [\ion{O}{i}] measurements could be introduced. The shaded region in Fig.\,\ref{fig:li_results} exemplifies this by showing the typical literature reference values of the Li feature, against those we observe here. Furthermore, since we show that stars with lower \teff\ are likely to have Li central positions red-wards of these literature reference values, it is possible that some velocity measurements may be more blue-shifted than previously measured. At a wavelength of 6300\,\AA, an introduced offset of +0.1\AA\ corresponds to a velocity shift of +4.8km/s. Hence, with this extreme example, the LVC centroids could be $\sim$-5\,km/s from the measurement values. Although the Li feature was not used for the TW Hya analysis in the previous example, for similar situations where the line was actually more blueshifted, it would help to reconcile the two suggested scenarios for the LVC emission, favouring the disk wind. 

\citet{simon_tracing_2016} noted proportionally fewer LVCs that are blue-shifted relative to the stellar photosphere than that of the study by \citet{hartigan_disk_1995}, even though the kinematic structure (i.e., the FWHM, presence of a BC or only NC) remained unchanged. The former study took the RVs calculated in \citet{pascucci_narrow_2015}, which used the Li line at a reference wavelength of 6707.83\,\AA\ for the majority of targets and the Ca I 6439.07\,\AA\ line for four targets. They reported a standard deviation of $\sim$0.8\,km/s between RVs measured using each absorption line, with differences of at most 4\,km/s from previous literature RV values\footnote{We have also checked in our data and the Ca I 6439.07\,\AA\ absorption line is always centred at 0\,km/s (within the errors) using the RVs we calculate.}. Our results suggest that outliers from the Li RV measurements could have shifted the [\ion{O}{i}] LVC centroids. Thus, if these LVCs were actually $\sim$5\,km/s more blueshifted, it may reduce the discrepancy found between these two studies. Indeed if fewer stars actually have [\ion{O}{i}] LVC velocities that are centred on the stellar photosphere rest frame, the notion of bound, low-density gas as the source of this emission is greatly reduced, with MHD or photoevaporative winds being the most likely instigator. 

The non-constant position of the Li feature, both across the stellar sample and from repeated observations of given targets, shows that this line should not be used exclusively for calibrating photocentric velocities at high spectral resolution. 


\section{Conclusions}

We report on a systematic study of the position of the Li absorption feature at $\sim$6708\,\AA\ in YSO spectra. We used the STAR-MELT package \citep{campbell-white_star-melt_2021} to automatically measure the stellar RV and fit models to 314 YSO spectra, resulting in 241 accurate measurements of the Li position, given the restrictions placed on the acceptable errors in RV and Gaussian fit centre (<2 km/s). 

We find that the mean wavelength of the Li absorption line is 6707.856\,\AA, with a standard error of 0.002\,\AA\ (0.09\,km/s) and a weighted standard deviation of 0.026\,\AA\ (1.16\,km/s). The maximum spread in measured values across all targets is 0.145\,\AA, or 6.5\,km/s. We discuss possible reasons for the overall mean and spread in values, which is likely due to the active nature and variability of YSOs.

We find a correlation between stellar \teff\ and the central position of the Li absorption, with higher \teff\ values having more blue-wards wavelengths ($r=-0.58$). Stars with higher \teff\ are known to have less overall Li abundance, which could in turn lead to the adjacent blends of e.g. Fe affecting the centroid of the line more significantly. Further potential sources of the spread in observed Li positions for YSOs may be from accretion-related RV shifts, star-spot coverage and both line/continuum veiling may have an effect on the measured parameters. We do not find any systematic trends in this respect, nor with the stellar \logg, but they are likely an additional source of scatter in the measures of the line centroid. Furthermore, we find that the temporal changes of the Li feature position may be varying on timescales shorter than the stellar rotation periods, but would require further repeated observations at short cadence to confirm this.

Finally, we report on the potential implications for photocentric or wavelength calibrations for kinematic studies of emission/absorption lines. From this empirical study of the position, we show that using only the Li feature may result in an introduction of $\sim$ +0.1\,\AA, which at 6300\,\AA\ results in a $\sim$ +5\,km/s offset. This means that studies of outflow-tracing forbidden line emission, such as [\ion{O}{i}], may have previously unknown systematic uncertainties. If these emission lines are actually more blueshifted than previously thought, the disk wind origin is much more likely than gas bound to the star/inner-disk.


\begin{acknowledgements}
We thank the referee, Eduardo Mart\'in, for their report that helped to improve this manuscript. We also thank Katia Biazzo for their useful input and discussion.
Funded by the European Union under the European Union’s Horizon Europe Research \& Innovation Programme 101039452 (WANDA). Views and opinions expressed are however those of the author(s) only and do not necessarily reflect those of the European Union or the European Research Council. Neither the European Union nor the granting authority can be held responsible for them. 
JCW and ASA were supported by the STFC grant number ST/S000399/1 ("The Planet-Disk Connection: Accretion, Disk Structure, and Planet Formation"). JCW also acknowledges funding from the SUPA Saltire grant and thanks J. Spyromilio for useful discussion. AF, BN, and MG  acknowledge support by the PRIN-INAF 2019 STRADE (Spectroscopically TRAcing the Disk dispersal Evolution) and by the Large Grant INAF YODA (YSOs Outflow, Disks and Accretion). AB acknowledges partial funding by the Deutsche Forschungsgemeinschaft Excellence Strategy - EXC 2094 - 390783311 and the ANID BASAL project FB210003.
This project has received funding from the European Research Council (ERC) under the European Union's Horizon 2020 research and innovation programme under grant agreement No 716155 (SACCRED). JFG was supported by funda\c c\~ao para a Ci\^encia e Tecnologia (FCT) through the research grants UIDB/04434/2020 and UIDP/04434/2020.

\end{acknowledgements}

%
%

\bibliography{bibliography.bib}

\appendix

\section{Li feature components and example Gaussian fits}
\label{app:fits_errors}

\begin{figure}[]
\centering

   \includegraphics[width=0.45\textwidth]{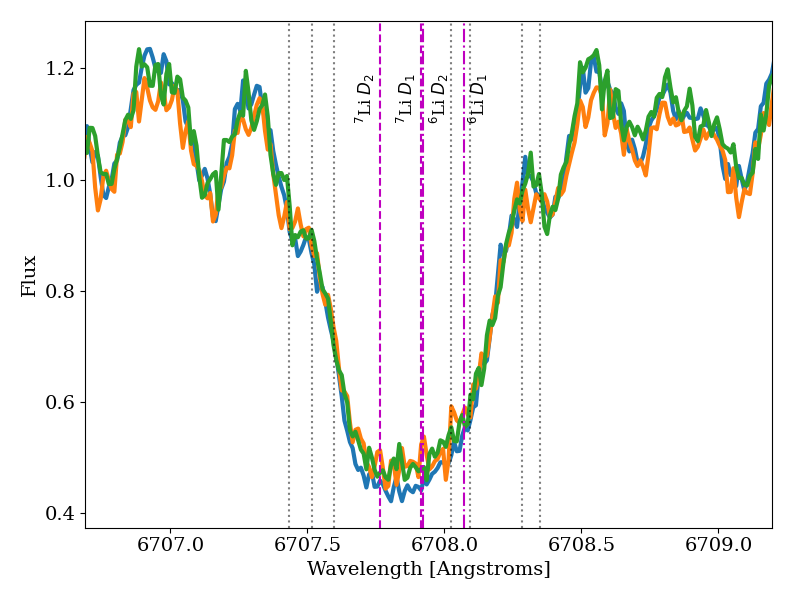}
\caption{ESPRESSO spectra of DM Tau around the Li 6708\,\AA\ feature. Positions of the Li isotope resonance $D$ lines are indicated by the magenta lines. Potential blends from elements with less abundance are marked by the black lines. From left to right, these are \ion{Fe}{i}, \ion{V}{i}, \ion{Cr}{i}, \ion{Si}{i}, \ion{V}{i}, \ion{Fe}{i} and \ion{Fe}{i} \citep{franciosini_gaia-eso_2022}. }
\label{fig:li_spec}
\end{figure}

\begin{figure}[]
\centering

   \includegraphics[width=0.49\textwidth]{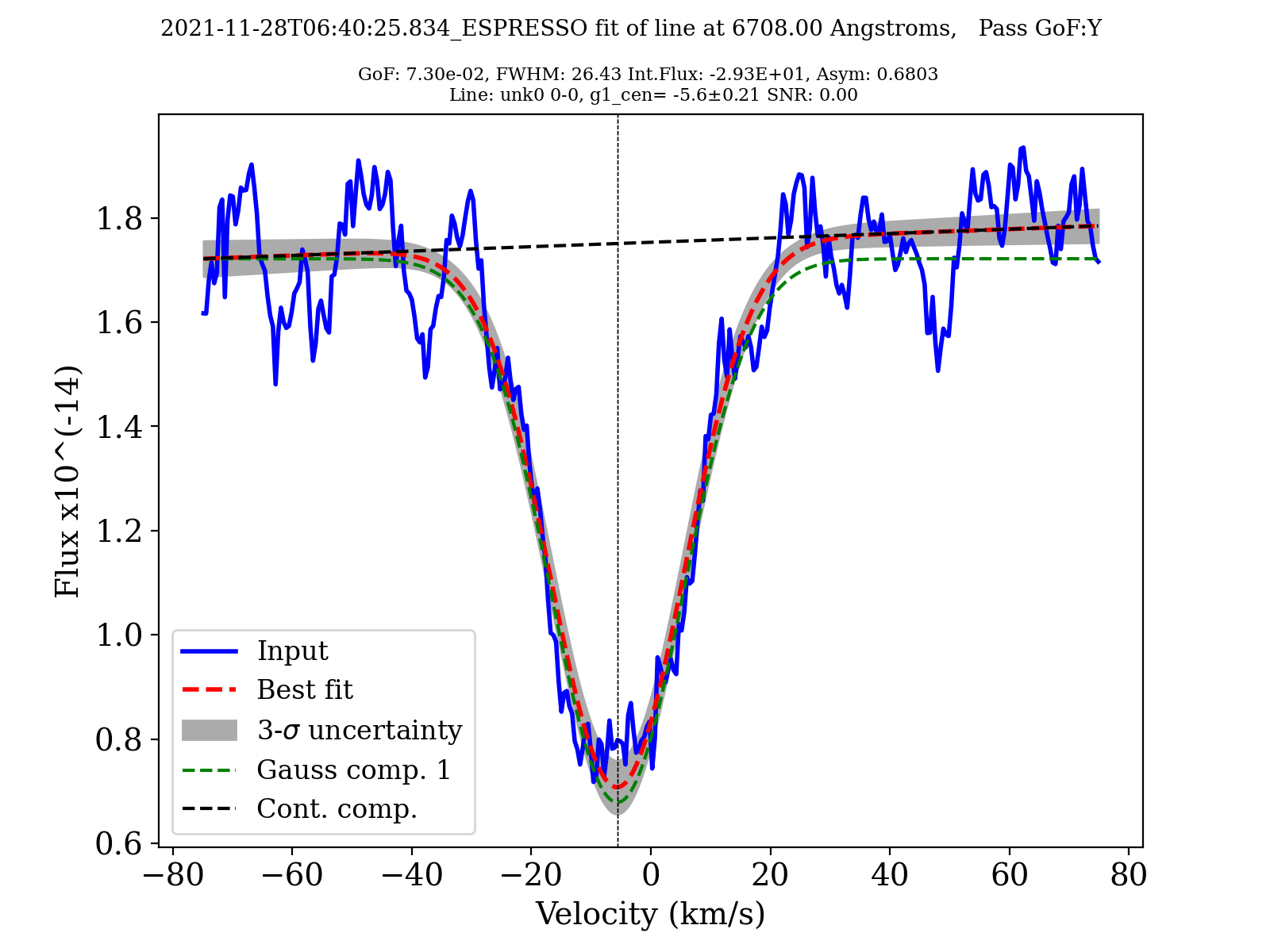}
      \includegraphics[width=0.49\textwidth]{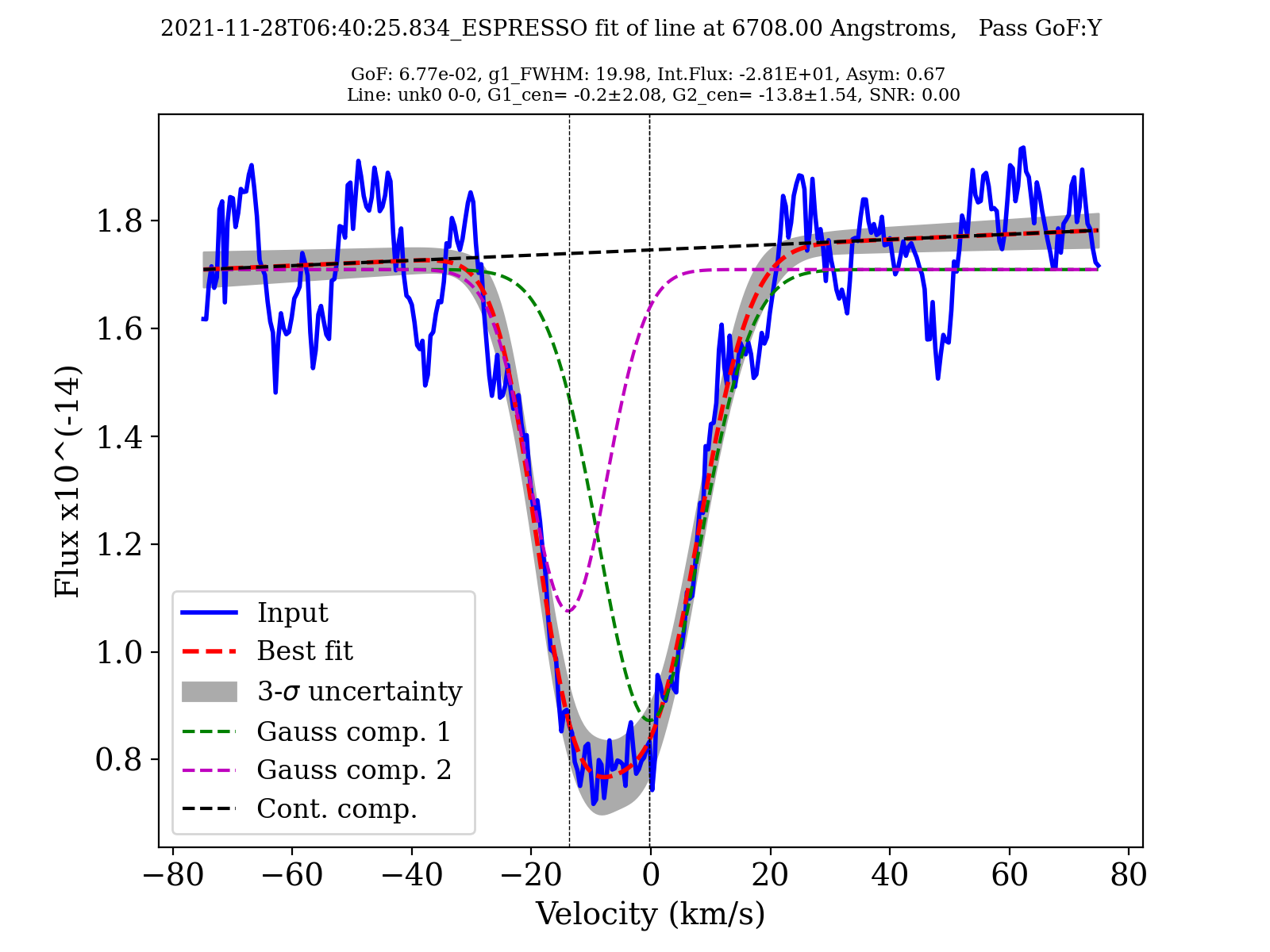}
  \caption{Example Gaussian fits to the ESPRESSO spectra of DM Tau for the Li absorption line. Central position of the Gaussian(s) are indicated by the vertical dashed line(s). }\label{fig:li_fits}
\end{figure}

\begin{figure*}[]
\centering
   \includegraphics[width=0.4\textwidth]{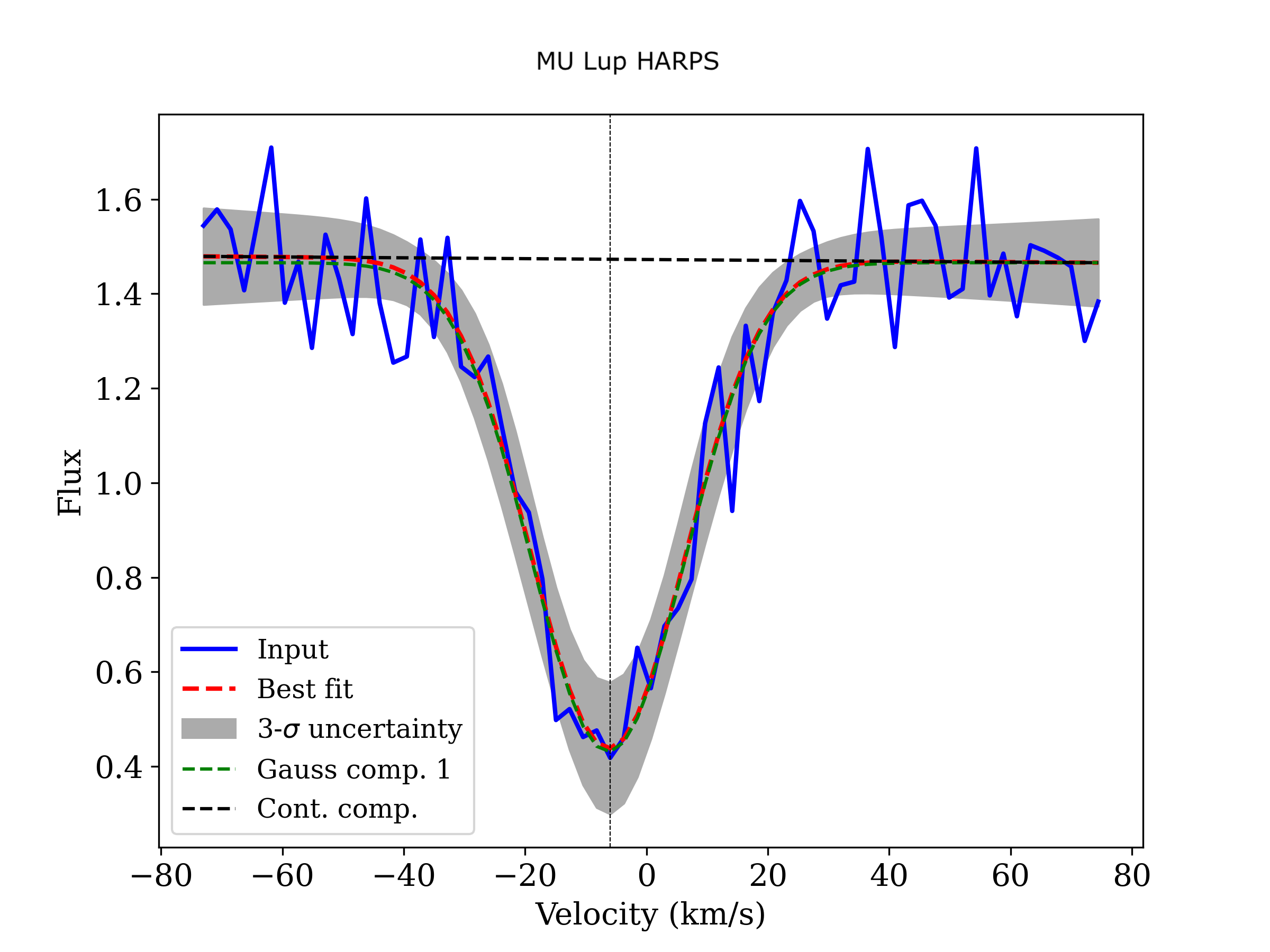}
   \includegraphics[width=0.4\textwidth]{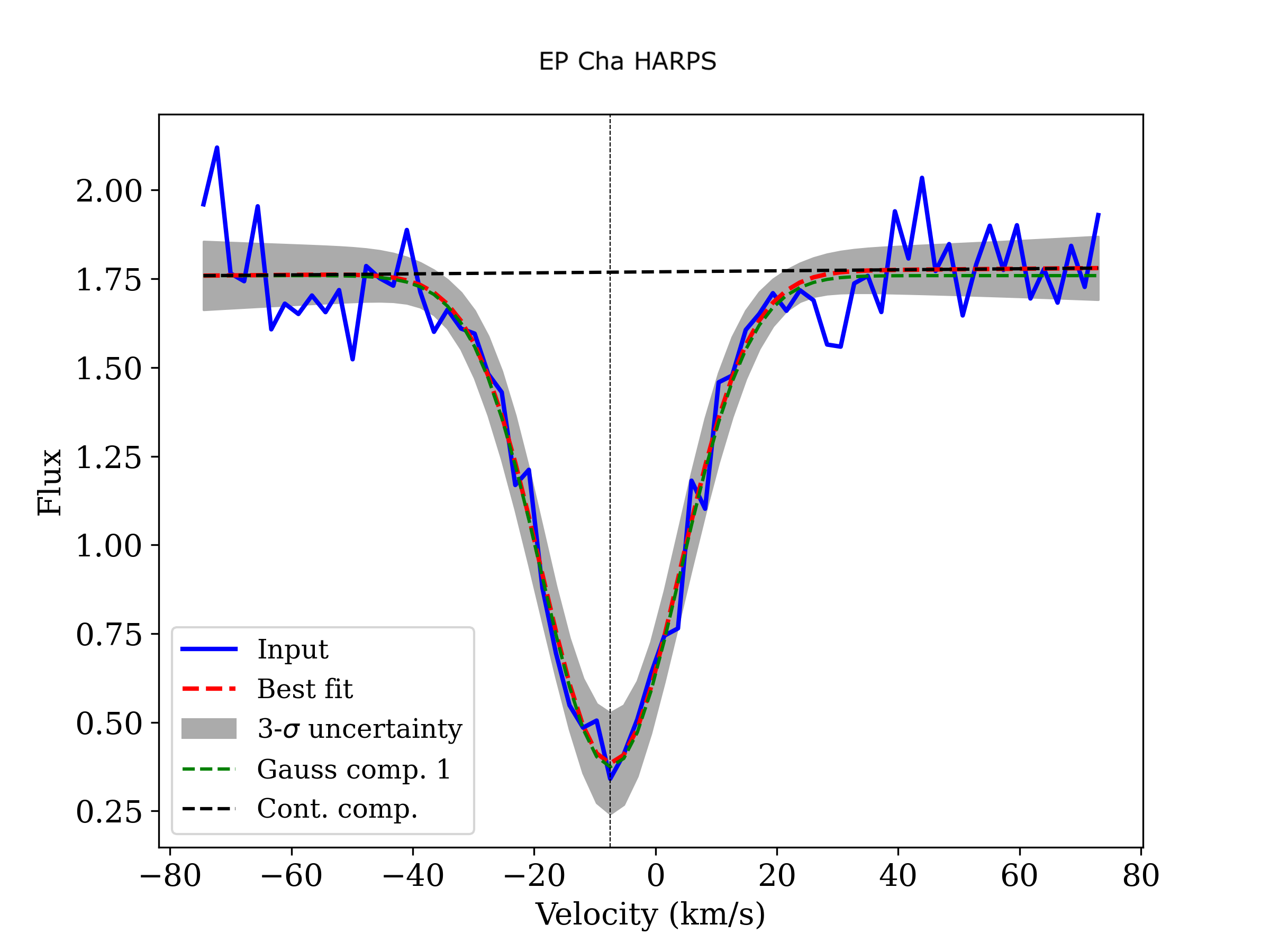}
   
    \includegraphics[width=0.4\textwidth]{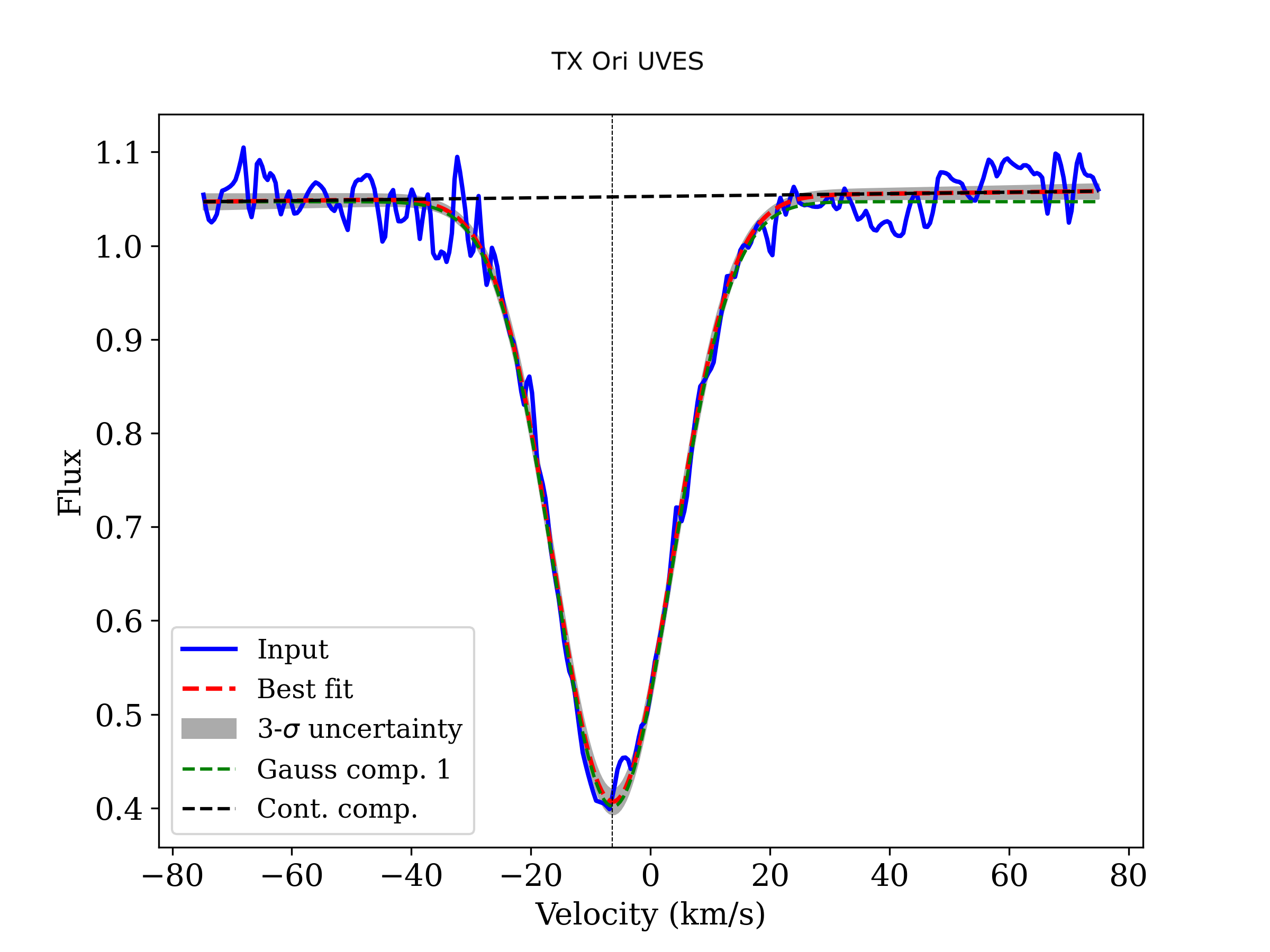}
    \includegraphics[width=0.4\textwidth]{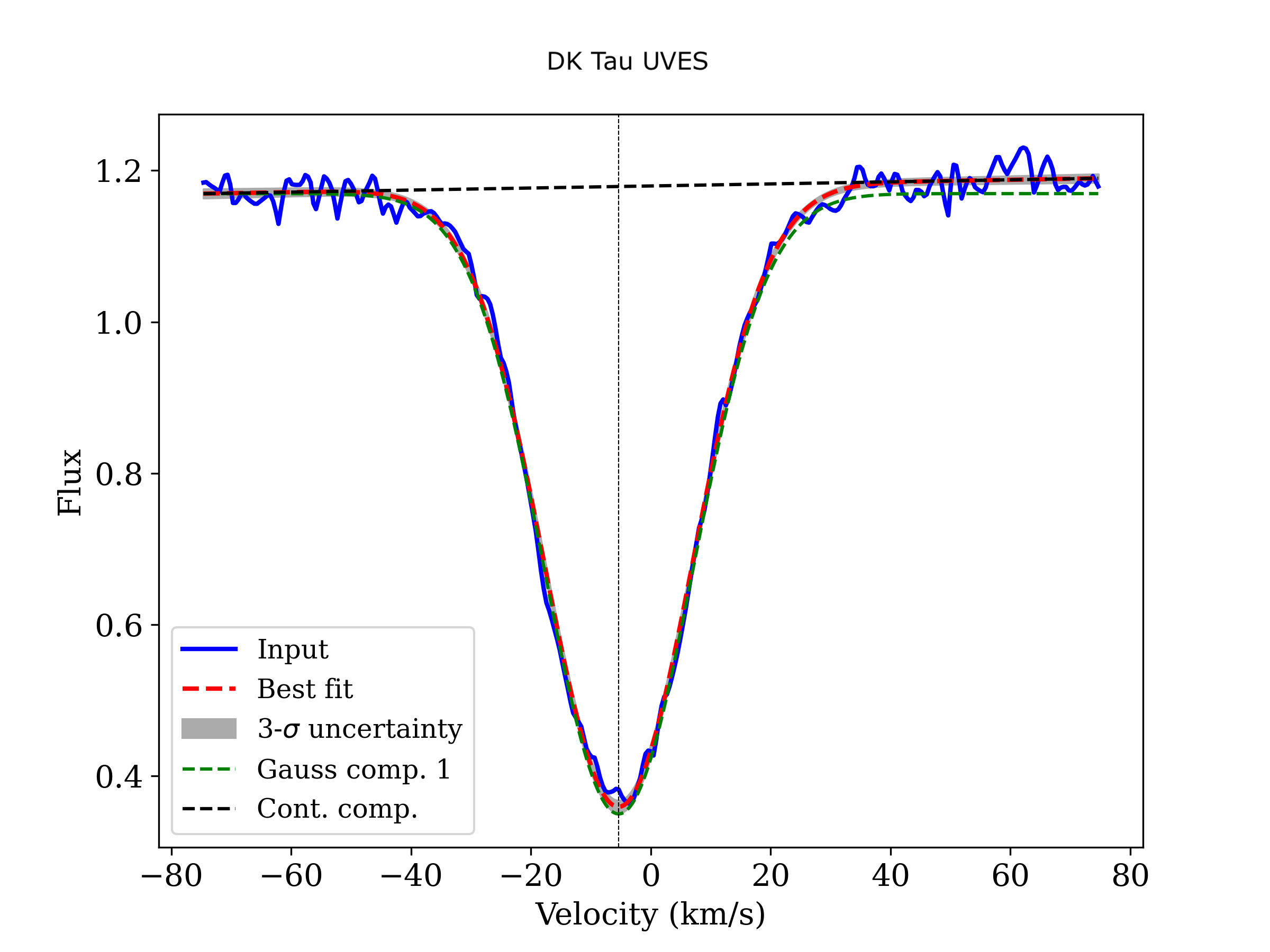}

    \includegraphics[width=0.4\textwidth]{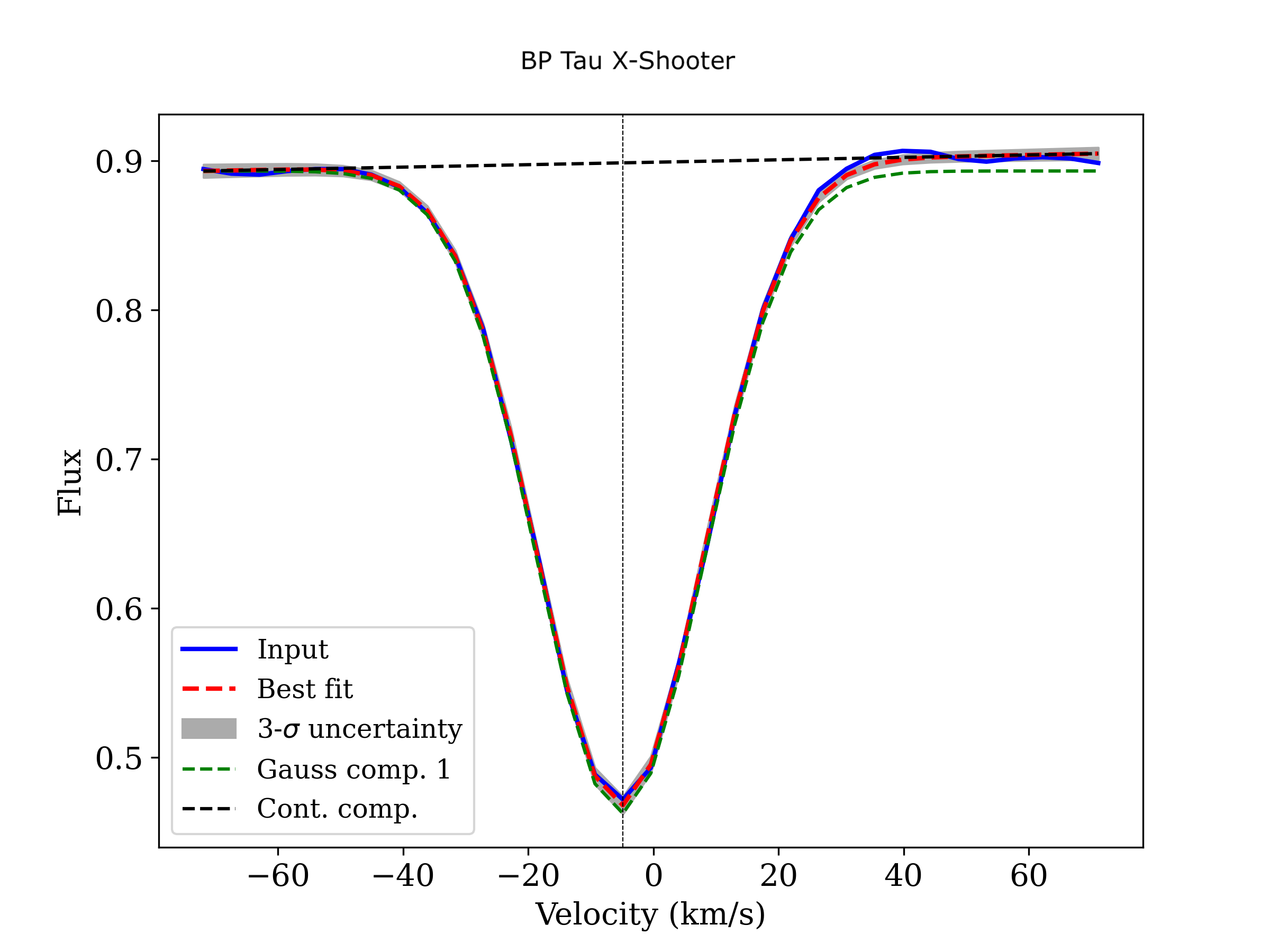}
    \includegraphics[width=0.4\textwidth]{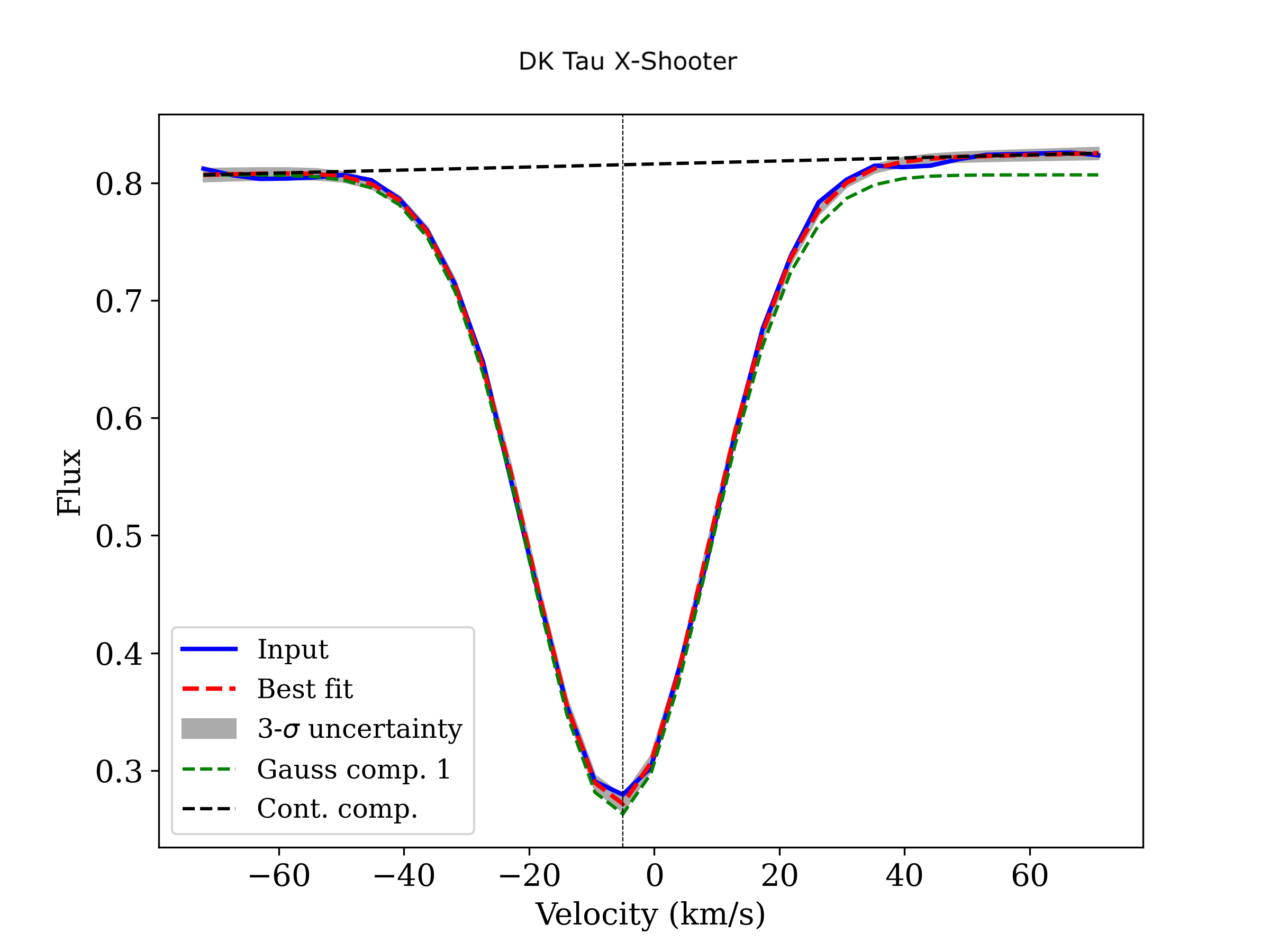}
    
    \includegraphics[width=0.4\textwidth]{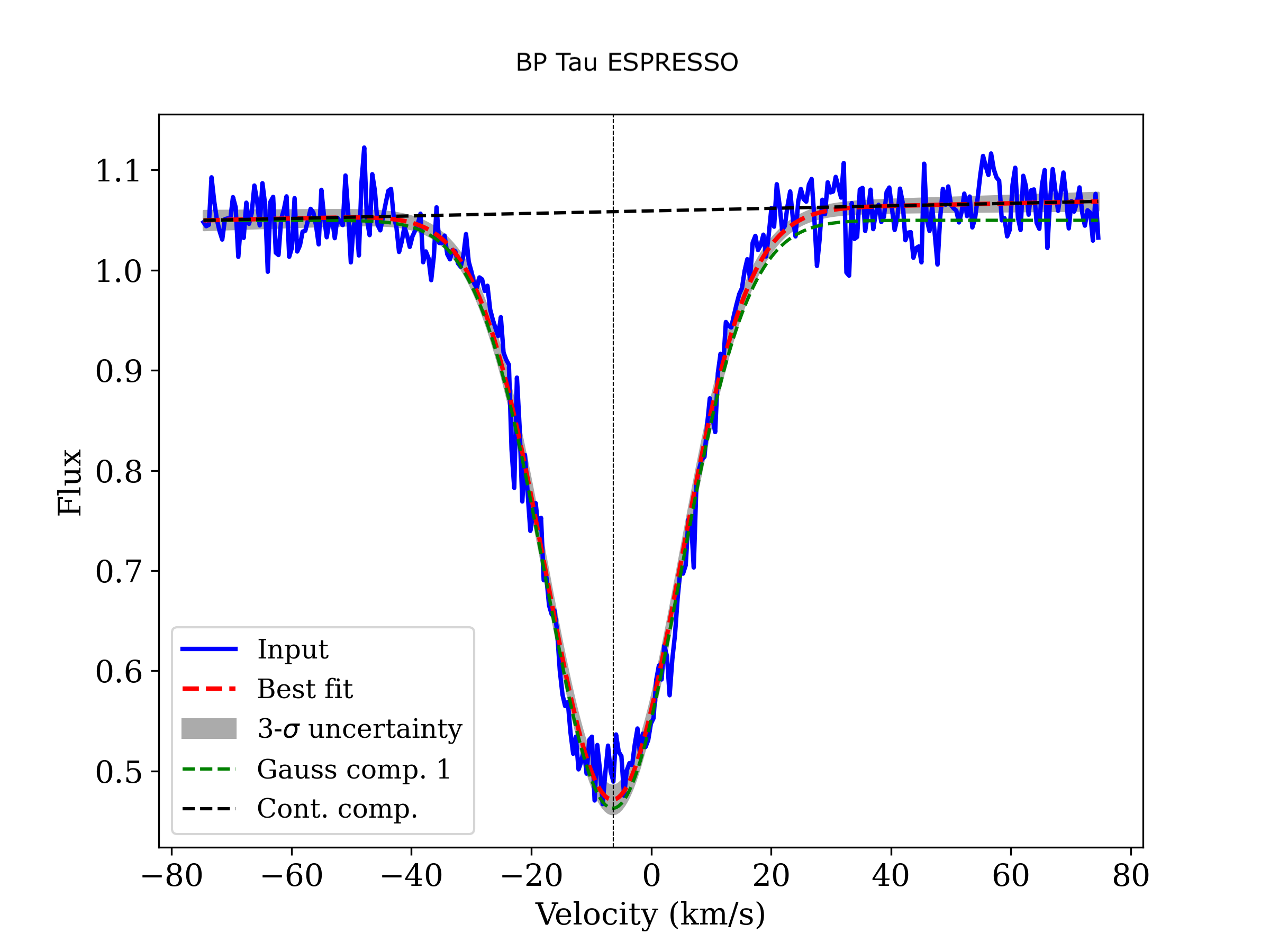}
    \includegraphics[width=0.4\textwidth]{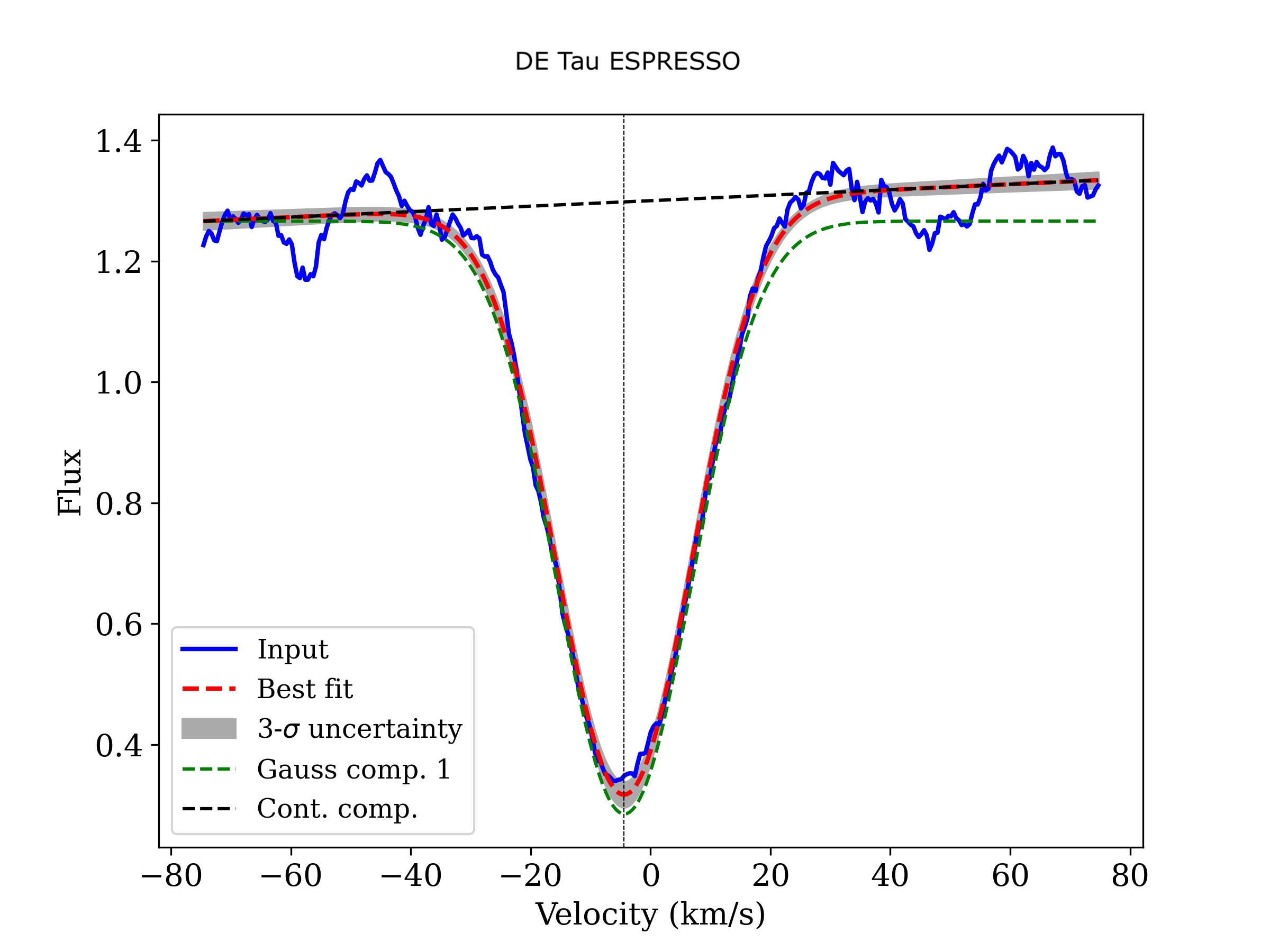}

  \caption{Further example Gaussian fits to the Li absorption line for the different instruments used. Central position of the Gaussian fits are indicated by the vertical dashed lines. Note that 0 km/s corresponds to a wavelength of 6708\,\AA, with Li centroid positions calculated with respect to this reference position. }\label{fig:li_inst_fits}
\end{figure*}

\begin{figure}[]
  \includegraphics[width=0.49\textwidth]{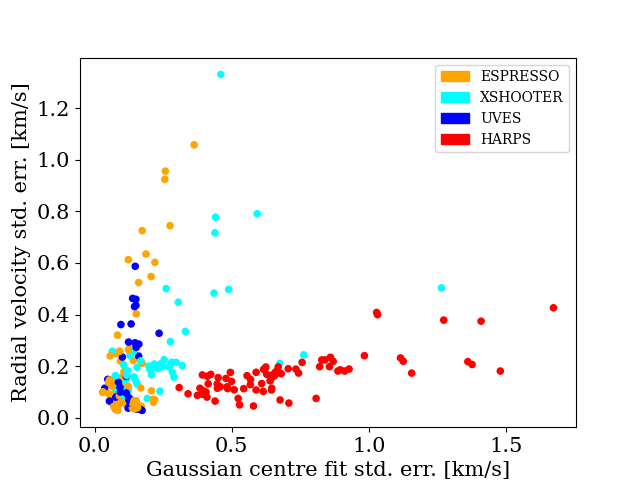}
  \caption{Calculated standard errors for the radial velocity measurements and the Gaussian centre position from the best-fit model to the absorption line.}
  \label{fig:errors}
\end{figure}

Figures showing Li Feature components (Fig.\,\ref{fig:li_spec}), example fitted Li profiles (Fig.\,\ref{fig:li_fits}), example fits for each instrument (Fig.\,\ref{fig:li_inst_fits}) and associated velocity errors (Fig.\,\ref{fig:errors}). For the measurement errors, from the automated fitting using STAR-MELT of the Li feature, typical standard errors for the RV or Gaussian centre measurements are $< 1$ km/s (see Fig.\,\ref{fig:errors}), we exclude any measurements with errors $> 2$ km/s. The Gaussian centre standard error is tightly correlated with the overall $\chi^2$ goodness-of-fit. We do not place a further restriction on the latter as it can be slightly higher due to noise in the continuum, yet still possess a well-fitted Gaussian component. Overall, the HARPS data had consistently low RV standard errors and the ESPRESSO and UVES measurements show the lowest Gaussian centre standard errors, due to both the higher S/N and resolution. 

\section{Further stellar properties comparison}
\label{app:logg_comp}

\begin{figure}[]
   \includegraphics[width=0.49\textwidth]{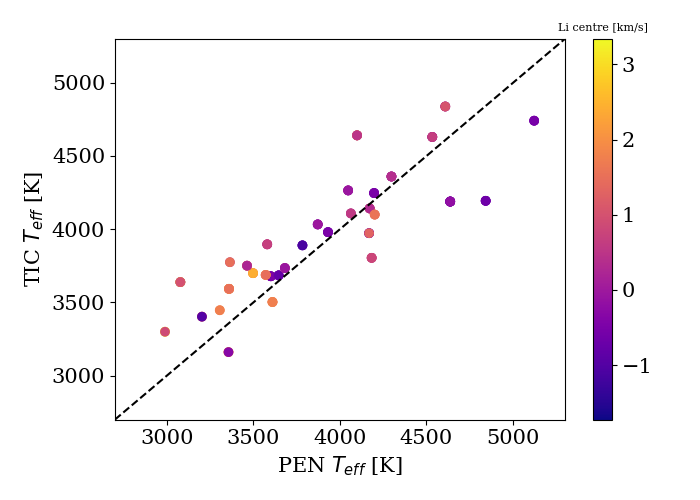}
\caption{Comparing the temperature differences from the ROTFIT derived spectra measurements of PENELLOPE to those from the Tess Input Catalogue. We find similar values of up to a few 100K lower in some spectrally derived temperatures, as in \citet{gangi_giarps_2022} and \citet{flores_effects_2022}. Points are coloured by the mean Li centroid position.}
\label{fig:teff_comp}

\end{figure}

\begin{figure}[]
   \includegraphics[width=0.49\textwidth]{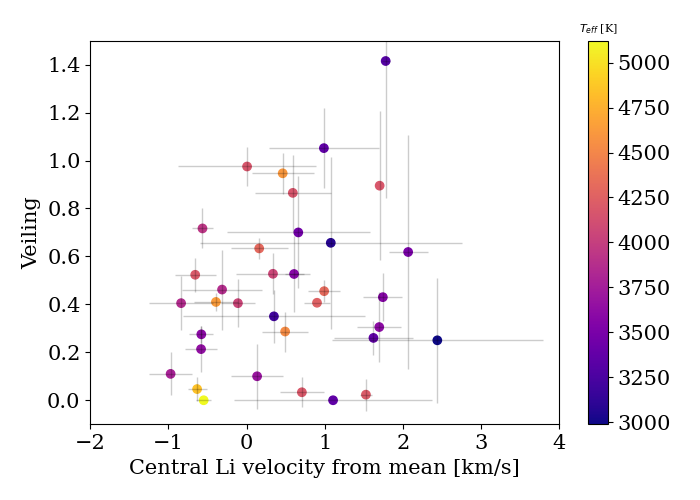}
\caption{Comparing the veiling measured with ROTFIT of the PENELLOPE spectra and the mean central velocity of the Li feature for each target. Veiling is measured within 50\,nm of the mean Li position. Points are coloured by the \teff\ measurement from ROTFIT. Pearson correlation coefficient of r=0.22, p=0.21 }
\label{fig:veil_comp}
\end{figure}

\begin{figure}[]
   \includegraphics[width=0.49\textwidth]{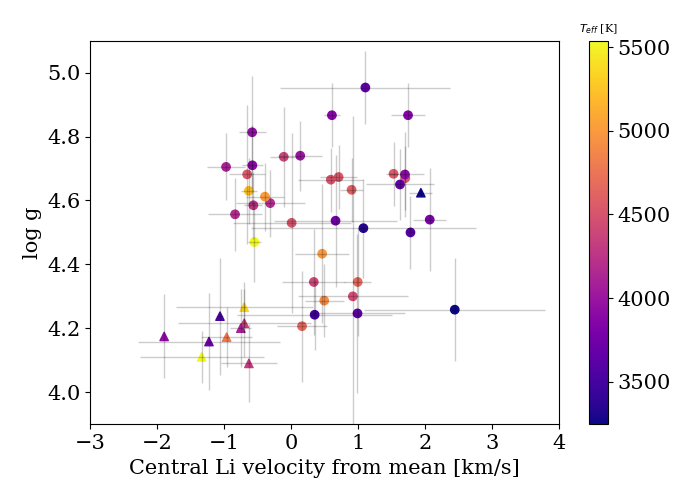}
\caption{Comparing the log g and the mean central velocity of the Li feature for each target, circles are ROTFIT measurements of the PENELLOPE spectra, triangles are from TIC. Points are coloured by the respective \teff\ measurement. Pearson correlation coefficient of r=0.35, p=0.02. }
\label{fig:logg_comp}
\end{figure}

Figures showing comparison of the \teff\ of the young star sample measured with ROTFIT for the PENELLOPE data and those from the Tess Input Catalogue (TIC) (Fig.\,\ref{fig:teff_comp}), and correlation plots for observed Li position versus veiling (Fig.\,\ref{fig:veil_comp}), and \logg\  (Fig.\,\ref{fig:logg_comp}) of the YSOs. Veiling and \logg\ measurements for the PENELLOPE sample are from ROTFIT, \logg\ values for the HARPS targets are from TIC. We find Pearson correlation coefficients for each that are three to six orders of magnitude less significant than the correlation between Li position and \teff, as reported in the main text.

\section{Individual observational results}
\label{app:individual_results}

\begin{figure*}[]
   \includegraphics[width=0.9\textwidth]{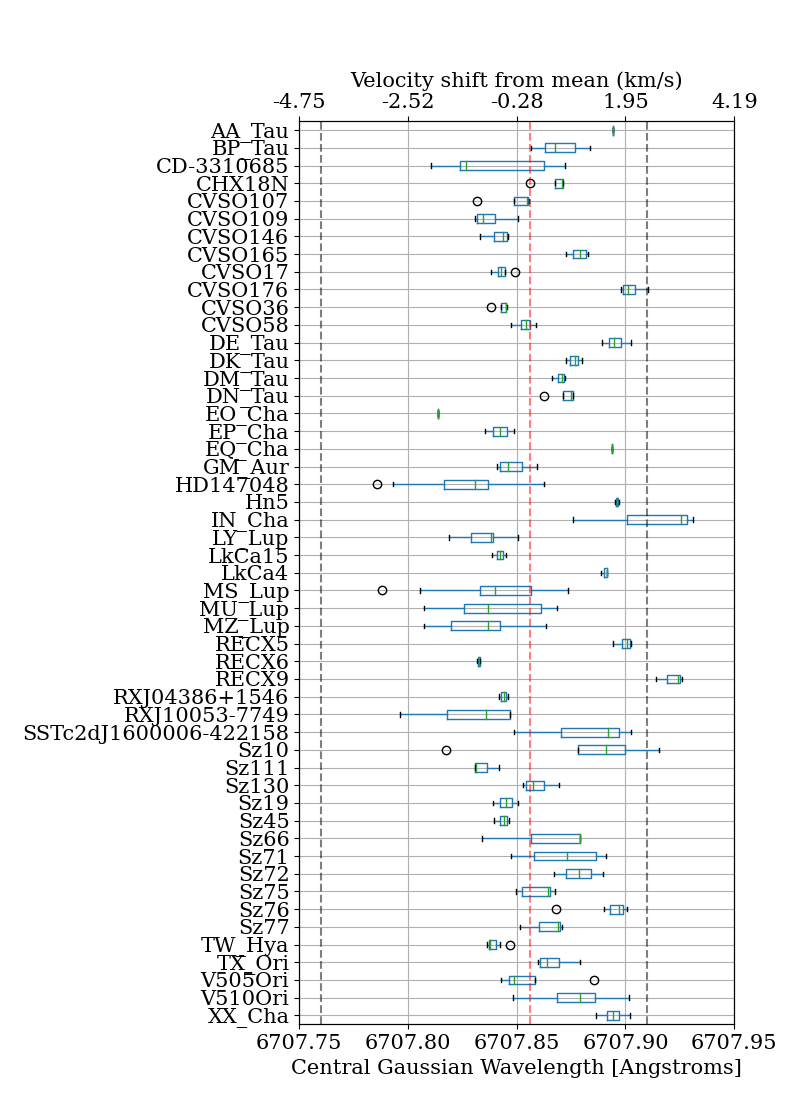}
\caption{Li measurement for each target. The mean position of the Li feature is indicated by the red dashed line. As with Fig.\,\ref{fig:li_results}, the boxes cover the inter-quartile range and whiskers 1.5 times this range. Outliers are indicated by the circles. Note that each of the PENELLOPE targets has at least one XSHOOTER observation in addition to their UVES/ESPRESSO observations. }\label{fig:li_all_bp}
\label{}
\end{figure*}

Here, we further the discussion on the spread of observed Li position from repeated observations of individual targets. Figure\,\ref{fig:li_all_bp} shows a box plot for the measured Li position of each target. For instrument differences between PENELLOPE observations using X-Shooter and UVES or ESPRESSO, TW Hya and GM Aur have the most repeated VLT observations and show good agreement between the X-Shooter and ESPRESSO measured Li positions. For targets with fewer observations but larger positional differences between the X-Shooter and ESPRESSO/UVES observations (see Tab\,\ref{tab:positions}), this may be due to instrument resolution differences (as previously described) or further inherent variability, since these dispersions do not exceed the range of those with more repeated observations.

\begin{figure*}[]
\centering
   \includegraphics[width=0.4\textwidth]{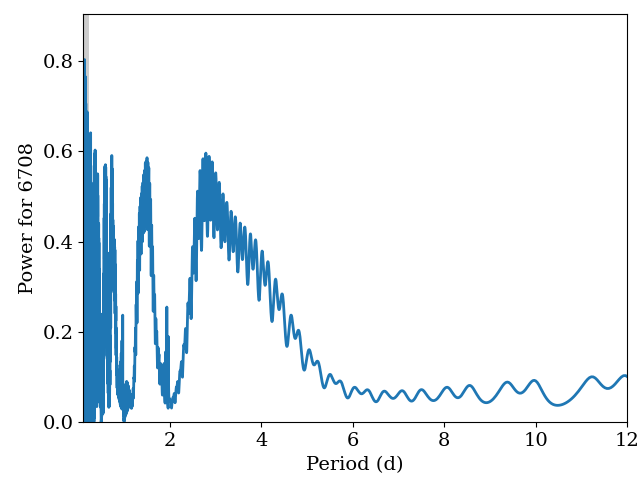}
   \includegraphics[width=0.4\textwidth]{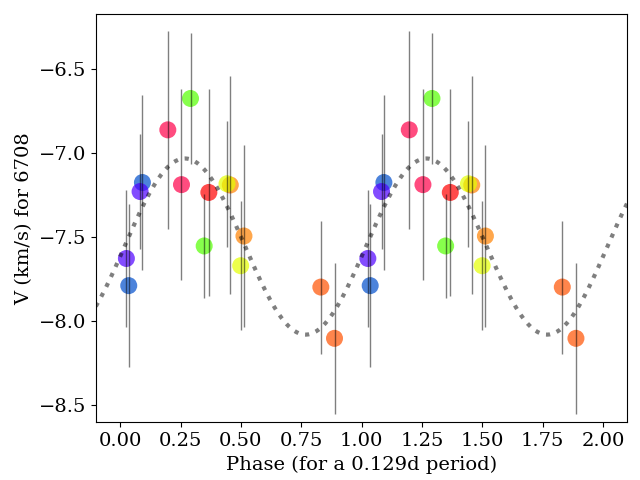}
   \includegraphics[width=0.4\textwidth]{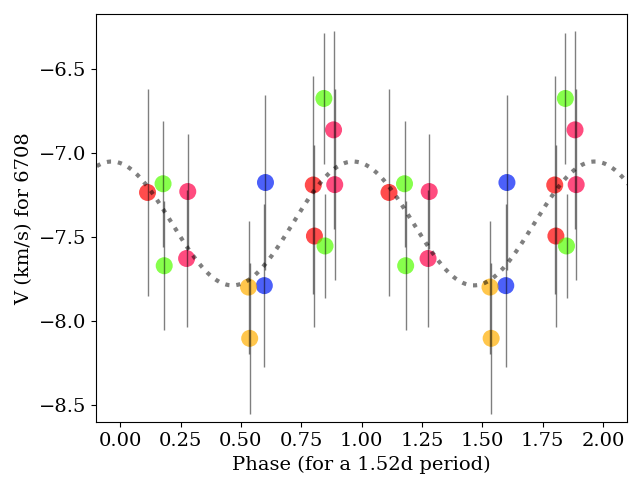}
   \includegraphics[width=0.4\textwidth]{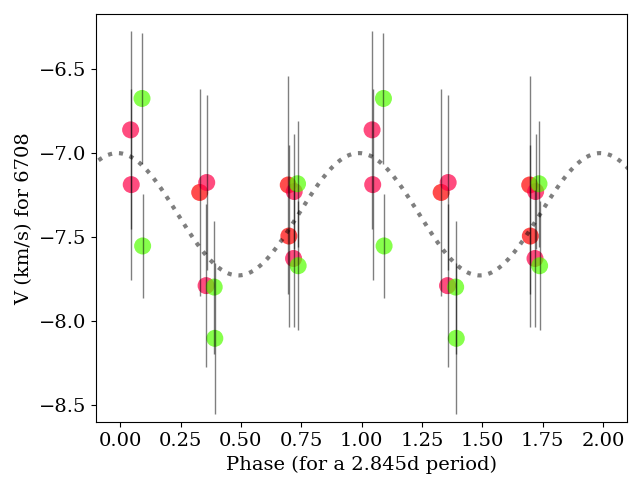}

\caption{LY Lup LSP and phase folded Li positions (procedure as described in \cite{campbell-white_star-melt_2021}. Note that the velocities indicated are those measured from the reference wavelength of 6708\,\AA, and are already corrected for stellar RV, hence the phase plots represent the relative shift in Li position. Points are coloured by observations corresponding to different phases. The photometric period of 2.84d from \citet{2012AcA....62...67K} is one of the peaks from the periodogram, however, shorter periods may be more representative of the observed temporal variations. Flase alarm probabilities are lower for the shorter periods ($\sim$ 20\%), however, still low significance due to the number of data points.}
\label{fig:ly_lup_phase}
\end{figure*}

\begin{figure*}[]
\centering
   \includegraphics[width=0.4\textwidth]{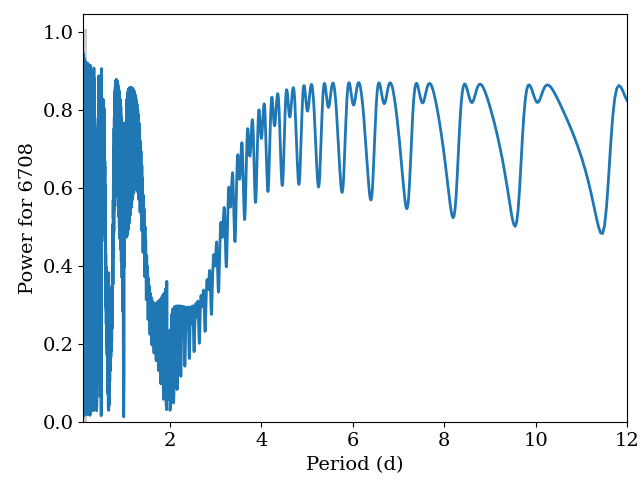}
   \includegraphics[width=0.4\textwidth]{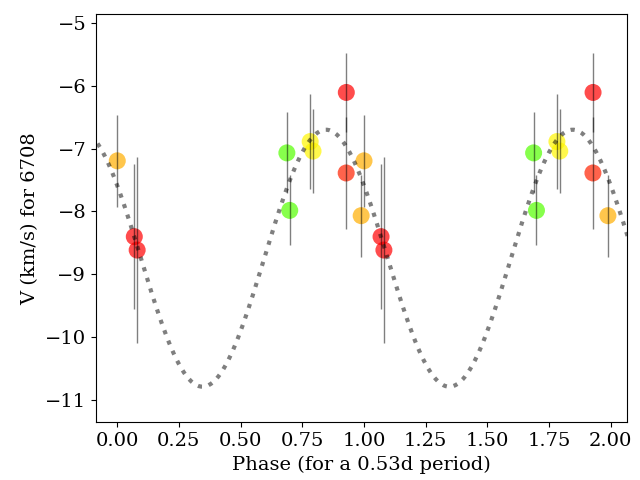}
\caption{As above for MZ Lup. Here, however, the photometric period of 4.47d from \citet{2012AcA....62...67K} is not clearly recovered in the periodogram. A high power period of 0.53d (false alarm probability of 10\%) is shown, which may suggest larger overall variations in position for the gaps in phase coverage. }
\label{fig:mz_lup_phase}
\end{figure*}

The stars LY Lup (15), HD147048 (16), MS Lup (16), and MZ Lup (10), had the most repeated observations with HARPS. We checked the time variation of the measured Li position versus the reported photometric rotational periods for these stars, finding that for LY Lup (see Fig.\,\ref{fig:ly_lup_phase}) and HD147048, the rotational period was recovered by the Lomb-Scargle periodogram \citep[LSP, for implementation details see][]{campbell-white_star-melt_2021}, however, shorter term variations had higher power, suggesting more rapid variations than those of the overall rotation. These periodicities are tentative (non-negligible false-alarm probabilities) due to the small number of data points, nevertheless, the phase plots illustrate well the spread in variations. For MS Lup and MZ Lup (see Fig.\,\ref{fig:mz_lup_phase}), the reported periods were not shown at significant power on the LSP, with multiple shorter periods less than a day found, again at low significance. Since in each case, we obtained repeated observations per night, and covered more than the rotational periods, these results suggest that the variability of the position of the Li centroid is perhaps on shorter timescales than typical rotation periods of YSOs.

\onecolumn
\LTcapwidth=\textwidth
\begin{longtable}{llrlrrrrr}

\caption{Summary of observations by instrument, mean RVs and Li shift (from the overall mean position) with standard deviations across observations as measured with STAR-MELT. For the ESPRESSO, UVES and X-Shooter spectra, \teff\ are also mean values measured across all PENELLOPE observations with ROTFIT, unless indicated by $*$. For these targets and the HARPS targets, \teff\ are from TIC \citep{stassun_revised_2019}.  }\label{tab:positions}\\
\toprule
      Target & Instrument    & No. Obs. &   Sp. Type & \teff\ & Mean RV & 1$\sigma$ RV & Mean Li shift & 1$\sigma$ Li shift \\
      &     &  &   & [K] & [km/s] & [km/s]  & [km/s]  & [km/s]  \\
\midrule
\midrule
\endfirsthead
\toprule
      Target & Instrument    & No. Obs. &   Sp. Type & \teff\ & Mean RV & 1$\sigma$ RV & Mean Li shift & 1$\sigma$ Li shift \\
      &     &  &   & [K] & [km/s] & [km/s]  & [km/s]  & [km/s]  \\
\midrule
\midrule
\endhead
\midrule
\multicolumn{9}{r}{{Continued on next page}} \\
\midrule
\endfoot

\bottomrule
\endlastfoot
AA Tau & XSHOOTER &   1 &    K7V & 4180 &    11.9 &        &         1.7 &            \\
\midrule
BP Tau & ESPRESSO &   2 &    K7V & 4172 &    15.8 &    0.6 &         0.2 &        0.2 \\
       & XSHOOTER &   3 &    & &    14.8 &    1.9 &         0.9 &        0.4 \\
\midrule
CD-3310685 & HARPS &   5 &   K3Ve & 4506 &    -1.6 &    0.6 &        -0.8 &        1.2 \\
\midrule
CHX18N & ESPRESSO &   3 &    K4V & 4534 &    15.3 &    1.3 &         0.7 &        0.0 \\
       & XSHOOTER &   2 &     & &    16.6 &    0.6 &         0.2 &        0.4 \\
\midrule
CVSO107 & UVES &   3 &    M0V & 3872 &    15.4 &    1.6 &        -0.4 &        0.6 \\
       & XSHOOTER &   1 &     & &    14.9 &        &        -0.0 &            \\
\midrule
CVSO109 & UVES &   3 &    M0V & 3861 &    16.5 &    0.5 &        -1.0 &        0.1 \\
       & XSHOOTER &   1 &    & &    16.2 &        &        -0.2 &            \\
\midrule
CVSO146 & ESPRESSO &   3 &    K6V & 4197 &    17.8 &    0.1 &        -0.5 &        0.1 \\
       & XSHOOTER &   1 &    & &    16.7 &        &        -1.0 &            \\
\midrule
CVSO165 & ESPRESSO &   3 &    K4V & 4314 &    26.7 &    0.4 &         0.9 &        0.2 \\
       & XSHOOTER &   1 &    & &    24.8 &        &         1.1 &            \\
\midrule
CVSO17 & UVES &   3 &    M1V & 3646 &    24.7 &    0.2 &        -0.7 &        0.1 \\
       & XSHOOTER &   1 &     & &    25.1 &        &        -0.3 &            \\
\midrule
CVSO176 & UVES &   3 &    M3V & 3498 &     9.2 &    1.4 &         2.0 &        0.1 \\
       & XSHOOTER &   1 &     & &     9.6 &        &         2.4 &            \\
\midrule
CVSO36 & UVES &   3 &    M1V & 3601 &    18.4 &    0.2 &        -0.6 &        0.2 \\
       & XSHOOTER &   1 &    & &    17.4 &        &        -0.5 &            \\
\midrule
CVSO58 & UVES &   3 &    K7V & 4048 &    21.5 &    1.6 &        -0.1 &        0.3 \\
       & XSHOOTER &   1 &     & &    21.5 &        &        -0.0 &            \\
\midrule
DE Tau & ESPRESSO &   3 &    M1V & 3571 &    14.3 &    0.0 &         1.8 &        0.2 \\
       & XSHOOTER &   1 &     & &    13.8 &        &         1.5 &            \\
\midrule
DK Tau & UVES &   2 &    K7V & 4241 &    14.4 &    1.8 &         1.0 &        0.1 \\
       & XSHOOTER &   1 &     & &    13.9 &        &         0.7 &            \\
\midrule
DM Tau & ESPRESSO &   3 &    M2V & 3580 &    18.7 &    0.1 &         0.7 &        0.1 \\
       & XSHOOTER &   1 &     & &    17.8 &        &         0.4 &            \\
\midrule
DN Tau & ESPRESSO &   3 &    K7V & 4183 &    16.9 &    0.4 &         0.8 &        0.0 \\
       & XSHOOTER &   1 &     & 4183 &    17.2 &        &         0.3 &            \\
\midrule
EO Cha & HARPS &   1 &     M0 & 3918 &    18.3 &        &        -1.9 &            \\
\midrule
EP Cha & HARPS &   2 &     K5 & 4308 &    15.3 &    0.2 &        -0.6 &        0.4 \\
\midrule
EQ Cha & HARPS &   1 &     M3 & 3546 &    21.8 &        &         1.7 &            \\
\midrule
GM Aur & ESPRESSO &   5 &    K4V & 4637 &    15.2 &    0.3 &        -0.3 &        0.2 \\
       & XSHOOTER &   4 &    & 4637 &    13.7 &    0.6 &        -0.4 &        0.4 \\
\midrule
HD147048 & HARPS &  16 &   G9IV & 5540 &     0.9 &    0.2 &        -1.3 &        0.9 \\
\midrule
Hn5 & UVES &   1 &    M4V & 3306 &    15.1 &        &         1.8 &            \\
       & XSHOOTER &   1 &    & 3306 &    15.6 &        &         1.7 &            \\
\midrule
IN Cha & UVES &   2 &    M4V & 2990 &    13.5 &    0.4 &         3.2 &        0.2 \\
       & XSHOOTER &   1 &    & 2990 &    15.1 &        &         0.9 &            \\
\midrule
LY Lup & HARPS &  15 &    K0e & 4791 &     4.7 &    0.3 &        -1.0 &        0.4 \\
\midrule
LkCa15 & ESPRESSO &   3 &    K4V & 4842 &    17.7 &    0.4 &        -0.6 &        0.1 \\
       & XSHOOTER &   1 &     & 4842 &    17.8 &        &        -0.8 &            \\
\midrule
LkCa4 & ESPRESSO &   3 &    K7V & 4201 &    16.9 &    0.7 &         1.5 &        0.1 \\
\midrule
MS Lup & HARPS &  16 &   G7IV & 5360 &     5.4 &    0.3 &        -0.7 &        1.0 \\
\midrule
MU Lup & HARPS &   8 &     K6 & 4354 &     2.6 &    0.5 &        -0.7 &        1.0 \\
\midrule
MZ Lup & HARPS &  10 &  G5IVe & 5105 &     2.3 &    0.3 &        -1.1 &        0.8 \\
\midrule
RECX5* & ESPRESSO &   3 &     M5 & 3245 &    17.2 &    0.1 &         2.0 &        0.1 \\
       & XSHOOTER &   1 &      & &    16.1 &        &         1.7 &            \\
\midrule
RECX6* & ESPRESSO &   2 &     M2 & 3522 &    18.1 &    0.1 &        -1.1 &        0.1 \\
\midrule
RXJ04386+1546 & ESPRESSO &   2 &    K2V & 5122 &    18.4 &    0.1 &        -0.5 &        0.1 \\
       & XSHOOTER &   1 &     & &    17.4 &        &        -0.7 &            \\
\midrule
RXJ10053-7749 & HARPS &   4 &    M1e & 3729 &    17.4 &    0.1 &        -1.2 &        1.1 \\
\midrule
SSTc2dJ1600006-422158 & UVES &   2 &  M3.5V & 3356 &     0.8 &    1.7 &         0.6 &        1.4 \\
       & XSHOOTER &   1 &  & &     3.2 &        &         2.1 &            \\
\midrule
Sz10 & ESPRESSO &   4 &    M4V & 3078 &    14.6 &    0.8 &         1.8 &        0.7 \\
       & XSHOOTER &   1 &    & &    18.8 &        &        -1.7 &            \\
\midrule
Sz111 & ESPRESSO &   2 &  M0.5V & 3784 &    -0.8 &    0.1 &        -1.1 &        0.0 \\
       & XSHOOTER &   1 &   & &    -0.7 &        &        -0.7 &            \\
\midrule
Sz130 & ESPRESSO &   3 &    M1V & 3682 &    -0.7 &    0.2 &        -0.0 &        0.2 \\
       & XSHOOTER &   1 &    & &     0.6 &        &         0.6 &            \\
\midrule
Sz45 & ESPRESSO &   3 &    K7V & 3931 &    14.4 &    0.2 &        -0.5 &        0.1 \\
       & XSHOOTER &   1 &     & &    13.8 &        &        -0.7 &            \\
\midrule
Sz66 & ESPRESSO &   2 &    M4V & 3203 &    -0.8 &    0.1 &         1.0 &        0.0 \\
       & XSHOOTER &   1 &     & &     1.1 &        &        -1.0 &            \\
\midrule
Sz71 & ESPRESSO &   3 &    M3V & 3463 &    -2.1 &    0.7 &         1.0 &        0.7 \\
       & XSHOOTER &   1 &     & &    -1.5 &        &        -0.4 &            \\
\midrule
Sz72 & ESPRESSO &   2 &    M3V & 3364 &    -2.5 &    0.4 &         1.0 &        0.7 \\
\midrule
Sz75 & ESPRESSO &   3 &    K4V & 4298 &    -2.6 &    0.4 &         0.4 &        0.1 \\
       & XSHOOTER &   2 &     & &    -3.3 &    0.6 &        -0.2 &        0.1 \\
\midrule
Sz76 & ESPRESSO &   4 &    M3V & 3359 &    -2.6 &    0.2 &         1.8 &        0.1 \\
       & XSHOOTER &   3 &     &  &    -3.1 &    0.6 &         1.3 &        0.7 \\
\midrule
Sz77 & ESPRESSO &   2 &   K6V & 4063 &    -2.0 &    1.1 &         0.6 &        0.1 \\
       & XSHOOTER &   1 &     &  &    -1.9 &        &        -0.2 &            \\
\midrule
TW Hya* & ESPRESSO &   5 &   K6Ve & 4097 &    12.6 &    0.1 &        -0.9 &        0.0 \\
       & XSHOOTER &   4 &    & &    11.4 &    0.4 &        -0.6 &        0.1 \\
\midrule
TX Ori & UVES &   3 &    K4V & 4608 &    30.0 &    0.9 &         0.3 &        0.2 \\
       & XSHOOTER &   1 &     & &    28.3 &        &         1.0 &            \\
\midrule
V505Ori & UVES &   3 &    K6V & 4168 &    31.0 &    1.1 &        -0.4 &        0.2 \\
       & XSHOOTER &   1 &     & &    31.7 &        &         1.3 &            \\
\midrule
V510Ori & ESPRESSO &   4 &    K7V & 4099 &    31.2 &    1.2 &         0.9 &        1.0 \\
       & XSHOOTER &   2 &     & &    31.9 &    0.1 &         1.0 &        0.4 \\
\midrule
XX Cha & UVES &   3 &    M1V & 3610 &    15.6 &    0.7 &         1.6 &        0.2 \\
       & XSHOOTER &   1 &     & &    14.6 &        &         2.0 &            \\
\end{longtable}

\end{document}